	\documentclass[twoside,journey]{IEEEtran}
	\usepackage{makecell}
	\usepackage{array}
	\usepackage{graphicx,amssymb,amsmath}
	\usepackage{multicol}
	\usepackage[noadjust]{cite} 
	\usepackage{setspace}
	\usepackage{subfigure}
	\usepackage{graphicx}
	\usepackage{float}
	\usepackage {url}
	\usepackage{stfloats}
	\usepackage{amsthm,pifont}
	\usepackage{flushend}
	\usepackage{cases,subeqnarray}
	\usepackage{bm,multirow,bigstrut}
	\usepackage{amsmath, amsthm, amssymb}
	\usepackage{textcomp}
	\usepackage{latexsym,bm}
	\usepackage{booktabs}
	\usepackage{xcolor}
	\usepackage{mathtools}
	\usepackage{dsfont}
	\usepackage{extarrows}
	\usepackage{epsfig}
	\usepackage{epsfig}
	\usepackage{epstopdf}
	\usepackage[noend]{algpseudocode}
	\usepackage{algorithmicx,algorithm}

	\theoremstyle{plain}

	\theoremstyle{plain}
	\newtheorem{rem}{Remark}

	\newtheorem{prop}{Proposition}
	\usepackage{amsmath}

	\IEEEoverridecommandlockouts
	
\begin{document}
	\title{Semantic Communications for Wireless Sensing: RIS-aided Encoding and Self-supervised Decoding}
	\author{Hongyang~Du, Jiacheng~Wang, Dusit~Niyato,~\IEEEmembership{Fellow,~IEEE}, Jiawen~Kang, Zehui~Xiong, Junshan~Zhang,~\IEEEmembership{Fellow,~IEEE}, and Xuemin~(Sherman)~Shen,~\IEEEmembership{Fellow,~IEEE}
	\thanks{H.~Du, J.~Wang and D. Niyato are with the School of Computer Science and Engineering, Nanyang Technological University, Singapore (e-mail: hongyang001@e.ntu.edu.sg, jcwang\_cq@foxmail.com, dniyato@ntu.edu.sg).}
	\thanks{J. Kang is with the School of Automation, Guangdong University of Technology, China. (e-mail: kavinkang@gdut.edu.cn)}
	\thanks{Z. Xiong is with the Pillar of Information Systems Technology and Design, Singapore University of Technology and Design, Singapore (e-mail: zehui\_xiong@sutd.edu.sg)}
	\thanks{J. Zhang is with the Department of Electrical and Computer Engineering, University of California Davis, USA (e-mail: jazh@ucdavis.edu)}
	\thanks{X. Shen is with the Department of Electrical and Computer Engineering, University of Waterloo, Canada (e-mail: sshen@uwaterloo.ca)}
	}
	\maketitle
	\vspace{-1cm}
	\begin{abstract}
	Semantic communications can reduce the resource consumption by transmitting task-related semantic information extracted from source messages. However, when the source messages are utilized for various tasks, e.g., wireless sensing data for localization and activities detection, semantic communication technique is difficult to be implemented because of the increased processing complexity. In this paper, we propose the {\textit{inverse semantic communications}} as a new paradigm. Instead of extracting semantic information from messages, we aim to encode the task-related source messages into a hyper-source message for data transmission or storage. Following this paradigm, we design an inverse semantic-aware wireless sensing framework with three algorithms for data sampling, reconfigurable intelligent surface (RIS)-aided encoding, and self-supervised decoding, respectively. Specifically, on the one hand, {\color{black}we propose a novel RIS hardware design} for encoding several signal spectrums into one {\textit{MetaSpectrum}}. To select the task-related signal spectrums for achieving efficient encoding, a semantic hash sampling method is introduced. On the other hand, we propose a self-supervised learning method for decoding the {\textit{MetaSpectrums}} to obtain the original signal spectrums. Using the sensing data collected from real-world, we show that our framework can reduce the data volume by $95\%$ compared to that before encoding, without affecting the accomplishment of sensing tasks. Moreover, compared with the typically used uniform sampling scheme, the proposed semantic hash sampling scheme can achieve $67\%$ lower mean squared error in recovering the sensing parameters. In addition, experiment results demonstrate that the amplitude response matrix of the RIS enables the encryption of the sensing data.
	\end{abstract}
	\IEEEpeerreviewmaketitle
	\section{Introduction}
	With the evolution of the next-generation Internet and the proliferation of wireless applications, the demand of network resources for data transmission, storage, and computation has been increasing rapidly.  {\color{black}Specifically, the advancement of technologies like extended reality and digital twins is driving the development of the Metaverse and Web 3.0 concepts~\cite{du2022attention}. As a result, there is an increasing demand for robust communication and computing support. To address the strict requirements of next-generation Internet applications, such as low latency, high reliability, and immersive experiences, semantic communication has been proposed as a key approach in the context of sixth-generation wireless communications~\cite{farshbafan2021common}. By transmitting only task-related semantic information extracted from source messages, semantic communications are believed to extend the conventional Shannon communication paradigm and bring higher quality of experience to users~\cite{seo2021semantics,yang2022semantic}.}
	
	{\color{black}While semantic communication techniques have demonstrated their significant effectiveness in processing source data across multiple modalities~\cite{xie2021task}, such as audio~\cite{weng2022deep}, image~\cite{du2023ai}, video~\cite{zhu2021video}, and text~\cite{xie2021deep},  the application of semantic communications to wireless sensing data processing remains a promising area with limited exploration thus far~\cite{yang2022semantic}.} The sensing data is important because that wireless signals are ubiquitous in our daily life, and can be used to accomplish various tasks requested by service providers. 
	Specifically, wireless signals not only help users access the Internet more efficiently, e.g., Metaverse, but also enable indoor positioning and activities detection more effectively. The wireless sensing data also facilitates the construction of virtual worlds such as digital twins. Unlike on-body sensor-based solutions~\cite{chen2019intelligent}, wireless sensing does not require the user to carry any devices and equipment, which is more practical and convenient. Additionally, the wireless sensing method is more robust than camera-based methods particularly in cases of occlusion or inadequate illumination, while causing fewer privacy issues.
	
\begin{figure*}[t]
	\centering
	\includegraphics[width=0.9\textwidth]{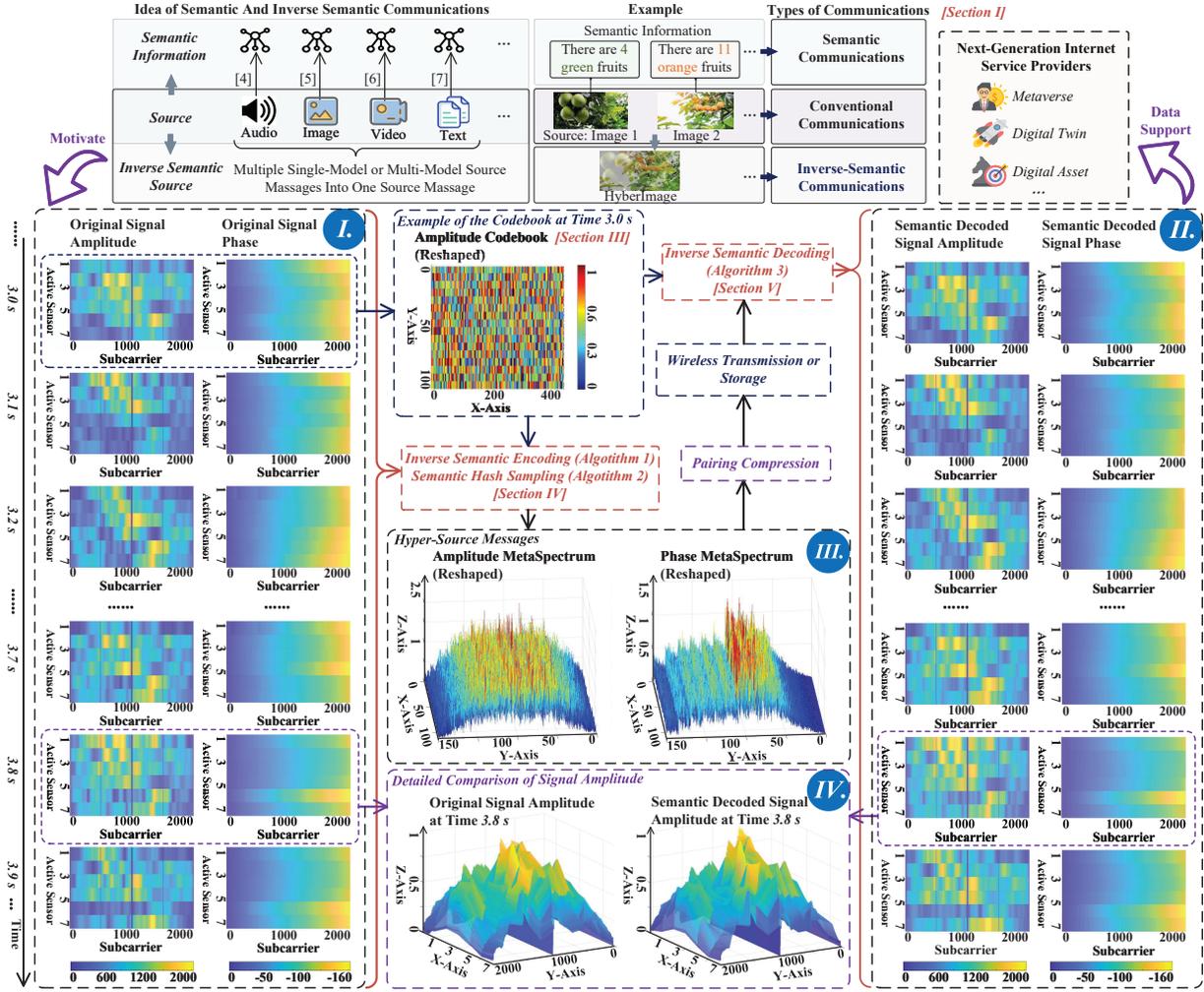} 
	\caption{The ideas of conventional, semantic, and inverse semantic communications. Motivated by the inverse semantic-aware communication, we propose an inverse semantic-aware encoding and decoding framework and show the results. Specifically, we select $10$ original signal amplitude/phase spectrum (Part I) by using our proposed {\bf{{Algorithm~\ref{Algorithm2}}}}, and encode them into one MetaSpectrum by using the RIS and {\bf{Algorithm~\ref{Algorithm1}}}. After wireless transmission, the reconstruction results (Part II) are obtained by decoding the MetaSpectrum using {\bf{Algorithm~\ref{Algorithm3}}}. The sensing data is collected by real experiments with an IEEE 802.11ax based test platform~\cite{gringoli2022ax}.}
	\label{res}
	\end{figure*}

	However, the wireless sensing technique has one major limitation. The transmission and storage of the sensing data, such as signal amplitude and phase spectrums, consumes a large number of resources~\cite{yang2022efficientfi}. In particular, the development of communication technologies such as multiple-input multiple-output and orthogonal frequency-division multiplexing (OFDM) improve the sensing resolution in the spatial and time-frequency domains, which, however, further increases the sensing data volume. Therefore, the semantic communication technique is expected to achieve efficient sensing data transmission or storage while achieving sensing tasks. {\color{black}This vision is more meaningful for applications that require long-term storage of sensing data, such as incremental learning for recognition \cite{ray2016survey}, healthcare services \cite{hassanalieragh2015health} and Internet-of-Things (IoT) systems and applications \cite{singh2020iot}.}
	The reason that semantic communications can ``exceed'' the Shannon limit is the ``impairment'' of the transmitted data, i.e., an effective semantic encoder extracts only task-independent semantic information from the source messages. However, a potential pitfall here is that the well-trained semantic encoding and decoding models for one specific task may fail when the source messages are needed to accomplish several different tasks. As shown at the top of Fig. \ref{res}, instead of transmitting an image, the semantic encoder can extract sentences describing the content of the image. This greatly reduces the number of bits that are required to be transmitted. However, semantic communications would not work well when the task is not only to know the type and number of fruits in the images, but also to know the spatial location. In this case, updated semantic models are required to be re-trained. In a word, semantic communications achieve efficiency transmission while introducing limitations. For the wireless sensing data, if we extract only the semantic information used for localization, the gesture detection task might not be accomplished.
	
	
	To fill this gap in semantic communications, we propose an inverse semantic-aware approach by treating the source messages as semantic information of a {\textit{hyper-source message}}. As shown in Fig.~\ref{res}, the ``inverse'' means that the processing of source messages is no longer to extract semantic information, but to combine multiple source messages (Part I) into one hyper-source message (Part III) for transmission or storage. Subsequently, by decoding, the semantic information of the hyper-source message (Part II), i.e., source messages, can be obtained to support multiple different tasks. Using the inverse semantic-aware approach, we reduce the data volume for transmission or storage, while avoiding the task limitations brought by semantic communications. For the wireless sensing, the source messages are signal amplitude and phase spectrums, and we call the hyper-source messages as amplitude and phase MetaSpectrums, respectively. We use the reconfigurable intelligent surface (RIS) to ensure efficient inverse semantic-aware encoding and decoding\footnote{{\color{black}Our scheme can alternatively be achieved by using active antennas and processors~\cite{gao2018compressive} to simulate the same signal processing as the RIS. However, higher hardware costs are introduced compared to the scheme using RIS.}}. With the RIS's superior ability to modulate signals, our scheme can be implemented effectively by modifying a small number of elements on the RIS without affecting the RIS-aided communications. {\textit{Unlike most RIS research works that consider only the phase response matrix of RIS, to the best of our knowledge, this is the first paper to make full use of the amplitude response matrix of RIS to help the system design for wireless sensing.}} The amplitude response matrix is not only used to reduce the sensing data volume significantly, but also to encrypt the sensing data because the amplitude response matrix is inevitable in the decoding process. Visual representation of the contributions of this paper is shown in Fig.~\ref{res}, which are summarized as follows:
{\color{black}
\begin{itemize}
	\item We propose a novel RIS hardware design following the paradigm of inverse semantic communications. The design features $L$-shaped active sensors placed behind transmissive elements in RIS, enabling inverse semantic-aware wireless sensing that significantly reduces the sensing data volume to 5\% of the original data volume, leading to improved efficiency and resource utilization.
	\item We develop the inverse semantic-aware encoding and decoding methods that leverage the amplitude response matrix of the RIS to embed prior knowledge in the sensing signals. The self-supervised learning-based decoding method requires no pre-training resource consumption and allows for the recovery of source messages to support multiple different tasks, overcoming the limitations encountered in semantic communications.
	\item We introduce an effective semantic hash sampling algorithm for selecting task-related sensing signal spectrums for decoding. Our approach achieves a mean squared error (MSE) between the ground truth and the 2D angles-of-arrival (AoA) estimation results that is 67\% lower than that of typically used uniform sampling schemes, and enhances security by utilizing the amplitude response matrix of the RIS for data encryption.
	\item We build an IEEE 802.11ax based test platform~\cite{gringoli2022ax} to collect real-world sensing data and perform experiments to demonstrate the effectiveness of our proposed framework.
\end{itemize}
}
	The remainder of the paper is organized as follows. In Section~\ref{Sre}, we review the related work in the literature. Section~\ref{Sre3} introduces the system model, which contains the novel RIS hardware and the sensing signal model. The inverse semantic-aware encoding and decoding methods are proposed in Section~\ref{S4ra} and Section~\ref{SS5}, respectively. Section~\ref{SS6} presents the experiment results. In Section~\ref{SF}, we present the conclusion and discuss some potential research directions.
	
	\section{Related Work}\label{Sre}
	In this section, we provide a brief review of three related techniques, i.e., wireless sensing, RIS, and spectral snapshot compressive imaging.
	\subsection{Wireless Sensing}
	{\color{black}Wireless signals, such as WiFi~\cite{niu2022rethinking}, have been widely used for sensing tasks ranging from large-scale intrusion detection to small-scale gesture recognition and breathing monitoring. With the rapid advancement of wireless sensing techniques, next-generation internet service providers can construct digital models of the physical world (for digital twin service) or conduct analysis of users' behaviors (for Metaverse services)~\cite{ramadan2020efficient,liu2019wireless,hassanalieragh2015health}. Wireless IoT devices collect the sensing data and use channel state information (CSI), which can be obtained as a sampled channel frequency response (CFR), for sensing tasks such as human activities detection~\cite{yue2020bodycompass} and passive localization~\cite{gao2022towards}. The CFR can be expressed as a complex matrix and decomposed into an amplitude spectrum and a phase spectrum for easy transmission and storage. With a three-dimensional multiple signal classification (3D-MUSIC) algorithm, the 3D spectrum can be obtained using the amplitude and phase spectra, which contain the time of flight (ToF) information. The obtained 3D spectrum can achieve several purposes, such as user localization~\cite{gong2021usability} or activity detection in the physical world. A challenge in this process is that the storage or transmission can cause excessive network resource consumption due to a large amount of sensing data.}
	
	\subsection{Reconfigurable Intelligent Surface}
	Significant developments in RIS-aided wireless communications have been witnessed over the past $3$ years, from hardware and algorithms design to deep integration with various technologies~\cite{10032267,yuan2022digital}. One of the most important application scenarios is to enhance wireless sensing~\cite{zhang2022toward}, such as indoor localization~\cite{zhang2021metalocalization} and direction-of-arrival estimation \cite{chen2022ris}. However, the existing methods typically aim to improve the sensing accuracy through signal enhancement by the RIS. The signal control capability of the RIS is not fully utilized, and most literature is limited in the study of reflective RIS that cannot achieve complete coverage. {\color{black}As our understanding of reconfigurable intelligent surface (RIS) hardware deepens, there has been a growing focus on transmissive and refractive RISs in the research community~\cite{tang2022transmissive,mu2021simultaneously,zhang2022dual}. Recently, novel RIS architectures such as simultaneously transmitting and reflecting (STAR) RIS~\cite{mu2021simultaneously} and intelligent omni-surface (IOS)~\cite{zhang2022dual} have been proposed to facilitate full-dimensional communications. We believe that the implementation of STAR RIS or IOS has the potential to significantly enhance wireless sensing capabilities. In addition to the direct benefit of improving sensing performance through signal enhancement, the ability to adjust the amplitude of transmissive signals can be leveraged as prior knowledge for the efficient compression of wireless sensing data. This paper will delve further into this concept and its implications for the field of wireless sensing.}

	\subsection{Spectral Snapshot Compressive Imaging}
	Capturing high dimension (HD) data is a long-term challenge in signal processing and related fields~\cite{yuan2021snapshot}. With theoretical guarantees, snapshot compressed imaging (SCI) uses two-dimensional (2D) detectors to capture HD, e.g., 3D, data in snapshot measurements using novel optical design. Then, reconstruction algorithms are applied to obtain the required HD data cubes~\cite{figueiredo2007gradient,meng2021self}. SCI has been used in many fields such as hyper-spectral imaging, video, holography, tomography, focal depth imaging, polarization imaging, and microscopy~\cite{wang2016adaptive}. However, there is no prior work discussing how to apply SCI to compressed sensing signals in the time dimension. The reason is that the highly dynamic nature of sensing signals brings difficulties to detector hardware design, coded aperture structure, and decompression algorithms. To fill this gap, in Section~\ref{CM}, we use the novel RIS hardware to perform one kind of special SCI to the sensing data. Using our proposed inverse semantic-aware encoding and decoding methods, the compression and self-supervised decompression of the sensing data can be achieved on time scale. Note that our design is different from compressive sensing (CS) methods in wireless sensing, and in fact can be used to further improve the performance of wireless CS systems.
	
	One primary objective of this study is to solve an important problem of overwhelming storage or transmission resources consumption in the wireless sensing. Inspired by the SCI system, we propose an encoding and decoding framework using the RIS to achieve inverse semantic-aware sensing, which significantly reduces the data volume and does not affect the accomplishment of various sensing tasks.

	\section{System Models}\label{Sre3}
	\begin{figure}[t]
	\centering
	\includegraphics[width=0.46\textwidth]{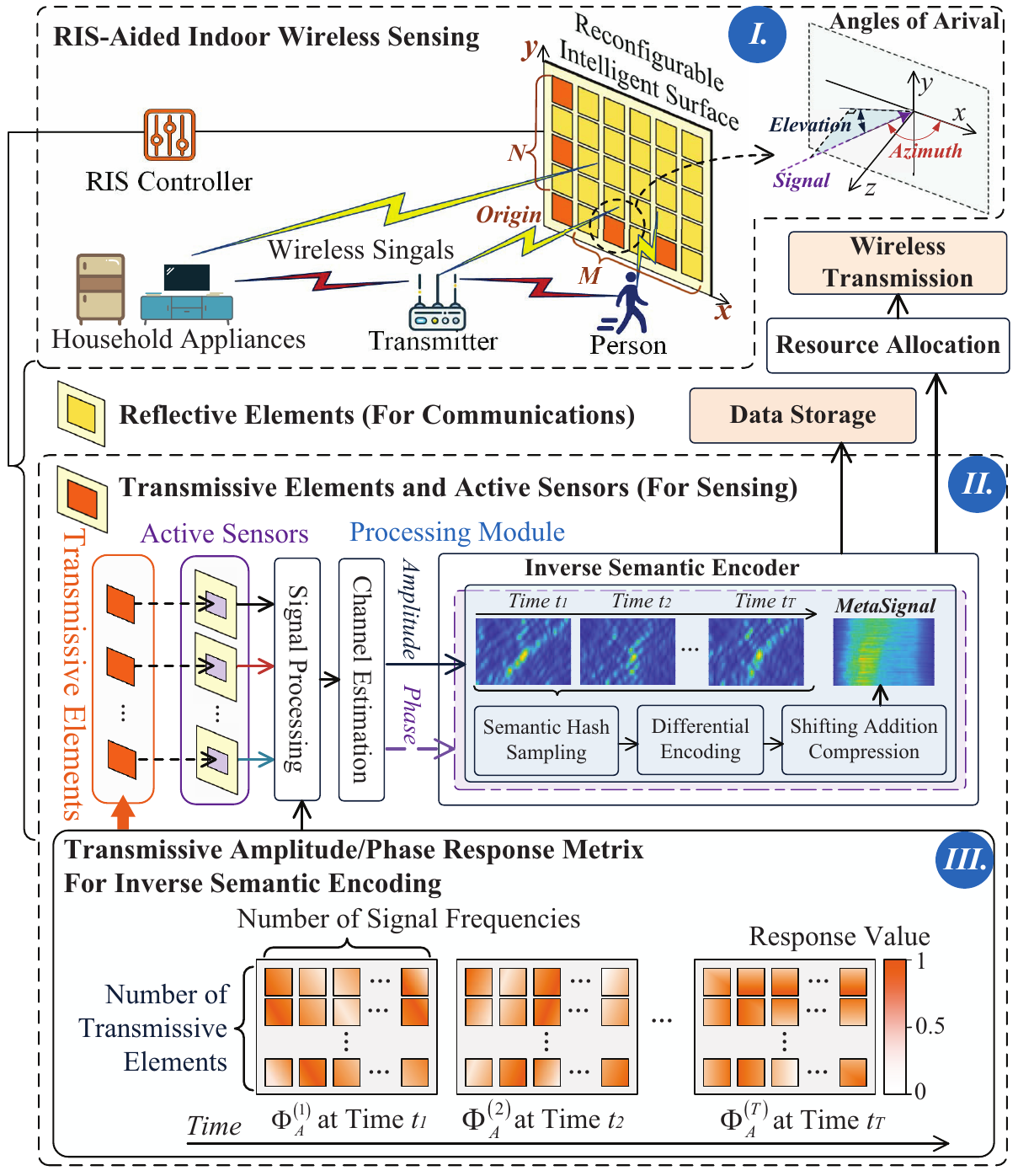} 
	\caption{The framework of the proposed inverse semantic-aware wireless sensing system.} 
	\label{img}
	\end{figure}
	Wireless signals contain user information such as activities and walking trajectories, and can preserve user privacy better than camera-based methods. Thus, mobile application SPs can use wireless signals to provide better services to users. For example, healthcare SPs can provide medical advice by analyzing the user's sleeping postures, and Metaverse SPs can customize virtual traveling scenes by positioning the users.
	To meet the needs of ubiquitous sensing data collection, we consider a 3D wireless indoor communications scenario as an example. As shown in Fig.~\ref{img} (Part I), a multi-antenna transmitter, e.g., IoT devices or WiFi router, transmits signals to multiple users with the help of an RIS. Different from the conventional scheme that uses RIS to improve sensing accuracy by enhancing signal strength, in this section, we propose a novel RIS hardware design to enable RIS with wireless sensing capability. Then, we analyze the mathematical formulas of the received sensing signals.
	
	
	\subsection{{\color{black} Novel Hardware of Reconfigurable Intelligent Surface}}
	To enable RIS to sense the environment, a widely used solution is to replace some reflecting elements on the RIS with active sensors, e.g., for channel estimation using CS~\cite{taha2021enabling}. Thus, a part of the RIS elements can switch between two operation modes, i.e., i) channel sensing mode that is used to estimate the channels, ii) reflection mode that reflects the signal. However, we can see that the RIS cannot assist communications in mode 1. We do not adopt directly the aforementioned solution since our goal is not merely to estimate the channel, but also constantly to sense the environment for the purposes of localizing and detecting user activities. {\color{black}To facilitate the sensing capabilities of the RIS without compromising its auxiliary communication functions, we initially integrate the RIS with a limited number of simultaneous transmitting and reflecting patches~\cite{mu2021simultaneously}, which are called transmissive elements in this paper for convenience. Specifically, as shown in Fig.~\ref{img} (Part II), the RIS is equipped with $L$-shaped $(M + N + 1)$ transmissive elements, and active sensors are strategically positioned behind them to receive signals modulated by the RIS.}
	\begin{rem}\label{L1}
	The reason for using the $L$-shaped array is that such a structure has more accurate 2D AoA estimation results than other structures, e.g., cross, linear, and rectangular arrays. This conclusion can be obtained by comparing the Cramer-Rao Bound metrics of different structures~\cite{hua1989shaped,zheng2021coupled}.
	\end{rem}
	
	Accordingly, the signal incident on the $q^{\rm th}$ transmissive element can be transmitted and reflected as~\cite{mu2021simultaneously}
	\begin{equation}
	{\beta_{i,q}}{{\rm exp}{\left( j{\delta_{i,q}}\right) }}, \qquad i \in \left\{ {T,R} \right\},
	\end{equation}
	where $i = T$ is for transmission coefficients and $i = R$ is for reflection coefficients. Note that, for each element, the responses of the RIS for transmission and reflection modes can be designed independently from each other~\cite{xu2021star}. In the following, we focus on the sensing function that only uses the transmitted signals. The reflection coefficients can be designed independently, which is outside the scope of this paper. Thus, after one path signal penetrates the $q^{\rm th}$ transmissive element on the RIS, the amplitude of the signal is multiplied by ${\beta_{T,q}}$, and the phase is added by ${\delta_{T,q}}$.
	
	{\color{black}A common assumption in much of the existing literature is that the amplitude and phase response of each element on the RIS remains constant throughout the signal bandwidth~\cite{wu2021intelligent,du2021millimeter}. While this assumption is generally acceptable for narrow bandwidths, it may become less accurate when dealing with multiple sub-carriers at varying frequencies within a broader range~\cite{wu2021intelligent,du2022performance}.} In our system model, we consider that the transmitter sends the wireless signals modulated by OFDM technology into $K$ sub-carriers\footnote{The OFDM is a widely used modulation method, which makes our analysis general. Moreover, OFDM can provide multi-carriers information, which is useful for signal parameters estimation.}. Because that $K$ might be large, e.g., $2048$ OFDM sub-carriers are used to transmit data in the IEEE 802.11ax protocol, we consider the practical case in which the element on the RIS has different responses to signals with different frequencies. Thus, the amplitude and phase response matrices of the $L$-shaped transmissive elements to $K$ sub-carriers at time $t$ can be expressed as
	\begin{equation}\label{CodeA}
	{\bf \Phi}_A^{\left(t\right)} =\!\! \left[\!\! {\begin{array}{*{20}{c}}
	{\beta _{{f_1}}^{[1,0]}}&\!\! \!\cdots\!\!\! &{\beta _{{f_1}}^{[M,0]}}&{\beta _{{f_1}}^{[0,0]}}&{\beta _{{f_1}}^{[0,1]}}& \!\!\!\cdots\!\!\! &{\beta _{{f_1}}^{[0,N]}} \\ 
	\!\!\!\vdots\!\! \!&\!\!\! \ddots\!\!\! &\!\!\! \vdots\!\!\! &\!\!\!\vdots\!\!\! &\!\!\!\vdots\!\!\! &\!\!\!\ddots\!\!\!& \!\!\!\vdots\!\!\!  \\ 
	{\beta _{{f_K}}^{[1,0]}}&\!\!\! \cdots\!\!\! &{\beta _{{f_K}}^{[M,0]}}&{\beta _{{f_K}}^{[0,0]}}&{\beta _{{f_K}}^{[0,1]}}& \!\!\!\cdots\!\!\! &{\beta _{{f_K}}^{[0,N]}} 
	\end{array}} \!\!\right]\!,
	\end{equation}
	and
	\begin{equation}\label{CodeP}
	{\bf \Phi}_P^{\left(t\right)} = \!\!\left[\!\! {\begin{array}{*{20}{c}}
	{\delta _{{f_1}}^{[1,0]}}&{\!\!\!\cdots\!\!\!}&{\delta _{{f_1}}^{[M,0]}}&{\delta _{{f_1}}^{[0,0]}}&{\delta _{{f_1}}^{[0,1]}}&{\!\!\!\cdots\!\!\!}&{\delta _{{f_1}}^{[0,N]}} \\ 
	{\!\!\!\vdots\!\!\!}&{\!\!\!\ddots\!\!\!}&{\!\!\!\vdots\!\!\!}&{\!\!\!\vdots\!\!\!}&{\!\!\!\vdots\!\!\!}&{\!\!\!\ddots\!\!\!}&{\!\!\!\vdots\!\!\!} \\ 
	{\delta _{{f_K}}^{[1,0]}}&{\!\!\!\cdots\!\!\!}&{\delta _{{f_K}}^{[M,0]}}&{\delta _{{f_K}}^{[0,0]}}&{\delta _{{f_K}}^{[0,1]}}&{\!\!\!\cdots\!\!\!}&{\delta _{{f_K}}^{[0,N]}} 
	\end{array}}\!\! \right],
	\end{equation}
	respectively, where $f_{i}$ denotes the frequency of the $i^{\rm th}$ sub-carrier, and $\left[x,y\right]$ denotes the location of the activate element. As shown in Fig.~\ref{img} (Part I), $0 \leqslant x \leqslant M$ and $0 \leqslant y \leqslant N$ indicate the locations of transmissive elements that are in the $X$-direction and $Y$-direction of the $L$-shape array, respectively.
	
	In the following, we analyze the sensing signal model, and propose the amplitude and phase response matrices design scheme. 
	
	\subsection{Sensing Signal Model}\label{S3B}
	To better understand the phase differences of the incident signals from different directions, we analyze the signals that impact the transmissive elements at different positions separately, i.e., the origin, the $X$-direction, and the $Y$-direction elements.
	At time $t$, the CFR of the multipath signals corresponding to the $k$-th OFDM sub-carrier obtained by the transmissive element located at the origin of the $L$-shaped array, can be expressed as
	\begin{align}\label{fml1}
	h_{f_k}^{\left[ {0,0} \right]}\left( t \right) = \sum\limits_{i = 1}^I {\alpha _i^{\left[ {0,0} \right]}{e^{ - j2\pi {f_k}{\tau _i}}}}, 
	\end{align}
	where $\left[ {0,0} \right]$ represents the origin point, ${\alpha _i}$ and ${\tau _i}$ are the complex signal amplitude attenuation and time delay of the $i$-th propagation path, respectively, ${f_k}$ is the frequency of the $k$-th sub-carrier, and $I$ is the total number of multipaths. Since different elements are placed at different positions in $L$-shaped array, the signal needs to travel different distances to arrive each transmissive element. Taking $h_{f_k}^{\left[ {0,0} \right]}$ as a reference, therefore, the CFR obtained by the $m$-th $X$-direction transmissive element can be expressed as
	\begin{align}\label{fml2}
	h_{f_k}^{\left[ {m,0} \right]}\left( t \right) = \sum\limits_{i = 1}^I {\alpha _i^{[m,0]}{e^{ - j2\pi {f_k}\left( {{\tau _i} + m\frac{{d\cos \left( {{\theta _i}} \right)\sin \left( {{\varphi _i}} \right)}}{c}} \right)}}},
	\end{align}
	where $\alpha _i^{[m]}$ is the signal amplitude attenuation, $d$ is the antenna spacing equals half wave length, ${\theta _i}$ and ${\varphi _i}$ represent the elevation angle and azimuth angle of the incident signal, respectively, as shown in Fig.~\ref{img} (Part I), and $c$ is the signal propagation speed in the air. Similarly, we can obtain the CFR of the $n$-th $Y$-direction transmissive element as 
	\begin{align}\label{fml3}
	h_{f_k}^{\left[ {0,n} \right]}\left( t \right) = \sum\limits_{i = 1}^I {\alpha _i^{[0,n]}{e^{ - j2\pi {f_k}\left( {{\tau _i} + n\frac{{d\sin \left( {{\theta _i}} \right)\sin \left( {{\varphi _i}} \right)}}{c}} \right)}}}.
	\end{align}
	Therefore, at time $t$, the overall CFR obtained by the $L$-shaped transmissive element array on the RIS can be expressed as a CFR matrix as follows:
	\begin{align}\label{fml6}
	&{{\bm{H}}_{xoy}^{\left( t\right) }} = \left[ {{{\bm{H}}_x^{\left( t\right) }}\;{{\bm{H}}_0^{\left( t\right) }}\;{{\bm{H}}_y^{\left( t\right) }}} \right]
	\notag\\&
	=\!\left[ {\underbrace {\begin{array}{*{20}{c}}
	{h_{{f_1}}^{\left[ {1,0} \right]}}& \!\!\!\cdots\!\!\!&{h_{{f_1}}^{\left[ {M,0} \right]}} \\ 
	\vdots & \!\!\!\ddots\!\!\!& \vdots  \\ 
	{h_{{f_K}}^{\left[ {1,0} \right]}}& \!\!\!\cdots\!\!\!&{h_{{f_K}}^{\left[ {M,0} \right]}} 
	\end{array}}_{{{\bm{H}}_x}}\underbrace {\begin{array}{*{20}{c}}
	{h_{{f_1}}^{\left[ {0,0} \right]}} \\ 
	\vdots  \\ 
	{h_{{f_K}}^{\left[ {0,0} \right]}} 
	\end{array}}_{{{\bm{H}}_0}}\underbrace {\begin{array}{*{20}{c}}
	{h_{{f_1}}^{\left[ {0,1} \right]}}&\!\!\!\cdots \!\!\!&{h_{{f_1}}^{\left[ {0,N} \right]}} \\ 
	\vdots & \!\!\!\ddots\!\!\!& \vdots  \\ 
	{h_{{f_K}}^{\left[ {0,1} \right]}}& \!\!\!\cdots\!\!\! &{h_{{f_K}}^{\left[ {0,N} \right]}} 
	\end{array}}_{{{\bm{H}}_y}}} \right].
	\end{align}
	%
	Note that ${{\bm{H}}_{xoy}^{\left( t\right) }}$ is the original sensing data that can support various sensing tasks. Because of the high available sampling frequency of the sensing device, e.g., $300$ times in one second~\cite{liu2019wireless}, and novel services that require long-term sensing, a large amount of ${{\bm{H}}_{xoy}^{\left( t\right) }}$ would be collected. To reduce the resources costed to store and transmit ${{\bm{H}}_{xoy}^{\left( t\right) }}$, we propose the inverse semantic-aware encoding and decoding methods in the following sections, respectively. 
	
	\section{Inverse Semantic-aware Encoding}\label{S4ra}
	In this section, we introduce the inverse semantic-aware RIS-aided encoding method to compress multiple signal spectrums into one. Two steps, i.e., differential encoding and shifting addition compression, are discussed. Moreover, we propose a semantic hash sampling method to select the task-related signal spectrum to record. 
	
	\subsection{Encoding Method}\label{CM}
	\begin{figure}[t]
	\centering
	\includegraphics[width=0.47\textwidth]{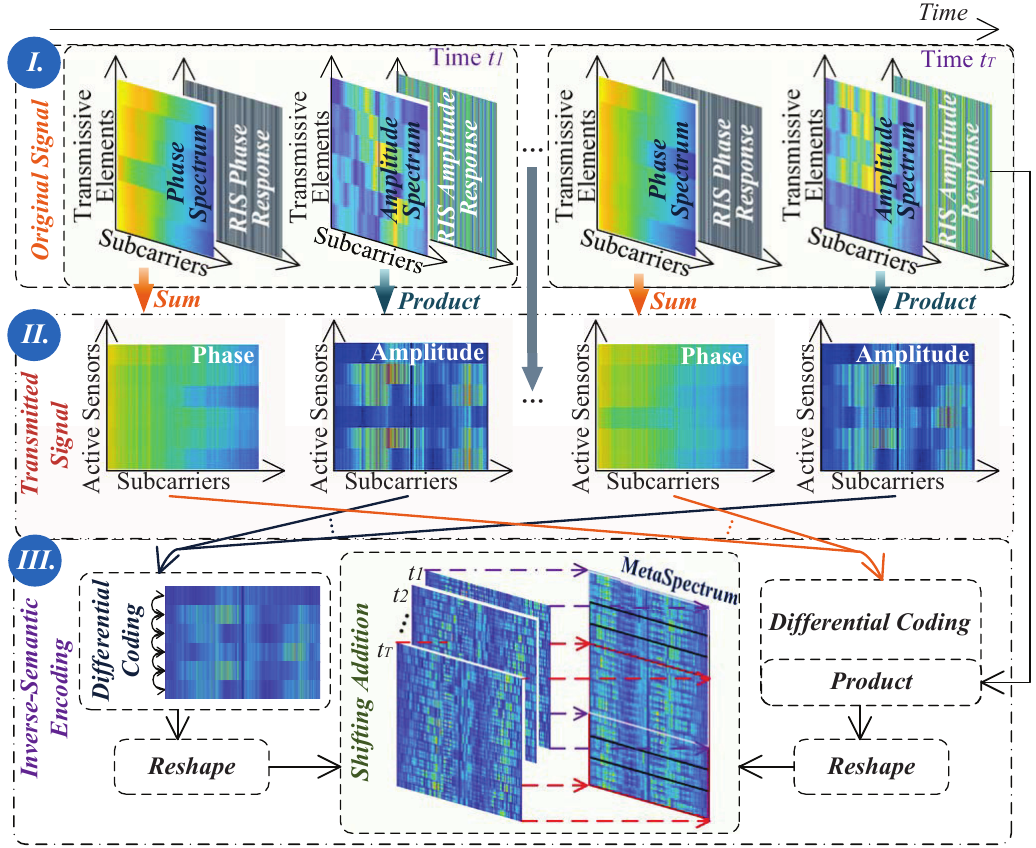} 
	\caption{The process of modulating the amplitude and phase spectrums through RIS, and then performing differential encoding and shifting addition.}
	\label{Compress}
	\end{figure}
	One can observe from \eqref{fml6} that every element in the CFR matrix is a complex number, which denotes the amplitude and phase of the CFR. Taking ${{\bf{H}}_x^{\left( t\right) }}$ as an example, it can be further decomposed into the amplitude and phase spectrums as
	\begin{align}\label{fml21}
	&	{{\bm{H}}_x^{\left( t\right) }} \to \left\{ {{{\bm{H}}_{x_a}^{\left( t\right) }},{{\bm{H}}_{x_p}^{\left( t\right) }}} \right\}
	\notag\\&
	=\!\! \left\{\! {\underbrace {\left[\!\!\! {\begin{array}{*{20}{c}}
	{\left\| {h_{f_1}^{\left[ {1,0} \right]}} \right\|}& \!\!\!\!\cdots\!\!\!\! &{\left\| {h_{f_1}^{\left[ {M,0} \right]}} \right\|} \\ 
	\!\!\!\!\vdots\!\!\!\! & \!\!\!\!\ddots\!\!\!\! & \!\!\!\!\vdots\!\!\!\!  \\ 
	{\left\| {h_{f_K}^{\left[ {1,0} \right]}} \right\|}& \!\!\!\!\cdots\!\!\!\! &{\left\| {h_{f_K}^{\left[ {M,0} \right]}} \right\|} 
	\end{array}} \!\!\!\right]}_{\textit{amplitude matrix}},\underbrace {\left[\!\!\! {\begin{array}{*{20}{c}}
	{\angle h_{f_1}^{\left[ {1,0} \right]}}& \!\!\!\!\cdots\!\!\!\! &{\angle h_{f_1}^{\left[ {M,0} \right]}} \\ 
	\!\!\!\!\vdots\!\!\!\! & \!\!\!\!\ddots\!\!\!\! & \!\!\!\!\vdots\!\!\!\!  \\ 
	{\angle h_{f_K}^{\left[ {0,1} \right]}}&\!\!\!\! \cdots\!\!\!\! &{\angle h_{f_K}^{\left[ {M,0} \right]}} 
	\end{array}} \!\!\!\right]}_{\textit{phase matrix}}} \!\right\}\!,
	\end{align}
	where $\cdot\!\to\!\cdot$ denotes the amplitude and phase extraction operation, $\left\{\cdot \right\}$ represents the set of matrices, $\left\|  \cdot  \right\|$ is the Euclidean norm operator, and $\angle h_{f_k}^{\left[ {m,0} \right]}$ denotes the signal phase of $h_{f_k}^{\left[ {m,0} \right]}$. Through the same way, ${{\bf{H}}_0}$ and ${{\bf{H}}_y}$ can be expressed as amplitude and phase spectrums, respectively. Hence, the CFR extracted from the $L$-shaped transmissive element array on the RIS can be expressed as 
	\begin{align}\label{fml22}
	{{\bf{H}}_{xoy}^{\left( t\right) }} = \left[ {{{\bf{H}}_x^{\left( t\right) }}{\ }{{\bf{H}}_0^{\left( t\right) }}{\ }{{\bf{H}}_y^{\left( t\right) }}} \right]\to\left\{{{\bf{H}}_A^{\left(t\right)}}, {{\bf{H}}_P^{\left(t\right)}}\right\},
	\end{align}
	where ${{\bf{H}}_A^{\left(t\right)}} = \left[{{\bm{H}}_{x_a}}\;{{\bm{H}}_{0_a}}\;{{\bm{H}}_{y _a}}\right]$ and ${{\bf{H}}_P^{\left(t\right)}} = \left[{{\bm{H}}_{x_p}}\;{{\bm{H}}_{0_p}}\;{\bm{H}}_{y_p}\right]$ denote the overall amplitude and phase at time $t$, respectively. As shown in Fig.~\ref{Compress} (Parts I and II), after being modulated by the transmissive elements, we can express $T$ received amplitude and phase spectrums by the $L$-shaped active sensor array in two sets as
	\begin{equation}\label{AmMask}
	{\bm{Y}}_A = \left\{ {\bm{H}}_A^{\left(1\right)} \circ {\bm{\Phi}} _A^{\left(1\right)},\ldots,{\bm{H}}_A^{\left(T\right)} \circ {\bm{\Phi}} _A^{\left(T\right)}\right\},
	\end{equation}
	and
	\begin{equation}\label{PhMask}
	{\bm{Y}}_P = \left\{{\bm{H}}_P^{\left(1\right)} + {\bm{\Phi}}_P^{\left(1\right)},\ldots,{\bm{H}}_P^{\left(T\right)} + {\bm{\Phi}}_P^{\left(T\right)}\right\},
	\end{equation}
	respectively, where $ \circ $ is the Hadamard product calculator, a.k.a., element-wise product.
	In the following, we encode the 3D data ${\bm{Y}}_A$ and ${\bm{Y}}_P$ onto 2D measurements, respectively. The encoding idea is inspired by the SCI system that compresses several optical spectrums of an object over multiple wavelengths into one spectrum, or several frames of a high-speed video into one frame. Specifically, the 3D data is first modulated by a coded aperture, and then spectrally dispersed by the dispersing element, and finally integrated across the spectral dimension to a 2D measurement. For the 3D sensing data ${\bm{Y}}_A$ and ${\bm{Y}}_P$, although the spectral dispersion process can be performed by low-power computing elements, there are several difficulties in adopting the compression scheme as in the SCI system:
	\begin{enumerate}
	\item[D1)] The fixed coded aperture in the SCI system is hard to be used in encoding signal spectrums that change dramatically on the time scale. However, the time-varying coded aperture scheme~\cite{qiao2020snapshot} increases the hardware cost and consume more storage space to record the patterns.
	\item[D2)] It is difficult for system designers to strike the balance between decoding performance and resource consumption. The inverse decoding problem is hard to be solved by traditional methods. {\color{black}The deep learning approach, such as convolutional neural networks, necessitates the use of extensive, well-labeled datasets and prolonged training periods~\cite{zhang2018ffdnet}, which restricts the frequency of aperture pattern updates.}
	\item[D3)] Signal spectrums are more sensitive than spectral images or video frames. We find from the experiments that the decoded sensing signal spectrums may lead to errors when performing some sensing tasks that are sensitive to the deviations in signal phase values, e.g., localization.
	\end{enumerate}
	To overcome the aforementioned difficulties, we rethink the SCI system from hardware design to software algorithms. For (D1) and (D2), we can observe from \eqref{AmMask} that the amplitude response matrix of the RIS has potential to perform a similar function as the coded aperture in the SCI system. It has been shown that the reconfiguration time for the RIS to change the response matrix is around $33$ ns~\cite{cui2020information}. Therefore, by changing the response matrix over time, the low-cost transmissive elements on the RIS can encode the sensing signals. In addition, the response values can be obtained from the hardware design parameters. This saves the storage resources to record a large number of original response values. Moreover, the amplitude and phase response values are discrete numbers, which can be determined by the number of coding bits. For example, $4$-bit coding bringse $16$ different available response values. Following that, we propose a self-supervision decoding algorithm for arbitrary RIS response matrices, which is discussed in Section~\ref{SS5}. Error negligible decoding results are achieved without pre-training resource consumption.

	To solve (D3), we compress the differential matrices of the amplitude and phase spectrums instead of the original spectrums to ensure the sensing performance. Unlike channel estimation, which focuses on accurately obtaining the CSI to better perform channel equalization, wireless sensing focuses on extracting information describing the physical environment from the CSI, e.g., 2D AoA and time of flight. This information is hidden in the value difference of amplitude and phase spectrums obtained by the sensors at different locations. For example, the phase difference between active sensors supports the signal AoA estimation. Another advantage to encode the differential spectrum is that the differential spectrum tends to be smoother than the original spectrum, due to the existence of correlation. This results in improved decoding performance. We show that real images can also benefit from the differential encoding in Fig.~\ref{realplot} using the dataset~\cite{perrin2020eyetrackuav2}.
	
	The differential encoding and shifting addition compression methods are presented in the following.

	\subsubsection{Differential Encoding}
	We first focus on the amplitude spectrum set ${\bm{Y}}_A$. Let ${\bm{Y}}_A\left\{i\right\}$ denote the $i^{\rm th}$ matrix in ${\bm{Y}}_A$. Each column in ${\bm{Y}}_A\left\{i\right\}$ represents the amplitude values of received signals at different frequencies by an active sensor, after amplitude modulation by the transmissive element on the RIS. Let ${\bm{Y}}_{A'}\left\{i\right\}$ denote the ${\bm{Y}}_A\left\{i\right\}$ after the differential encoding. Specifically, we let the $j^{\rm th}$ column in ${\bm{Y}}_{A'}\left\{i\right\}$ store the difference values of the $j^{\rm th}$ column and $\left(j-1\right) ^{\rm th}$ column in ${\bm{Y}}_A\left\{i\right\}$ as 
	\begin{equation}\label{q1}
	{{\bm{Y}}_{A'}}\left\{ i \right\}\left[ {j,:} \right] = {{\bm{Y}}_A}\left\{ i \right\}\left[ {j ,:} \right] - {{\bm{Y}}_A}\left\{ i \right\}\left[ {j- 1,:} \right],
	\end{equation}
	where $j = 2,\ldots,L$. The first columns in ${\bm{Y}}_{A'}\left\{i\right\}$ and ${\bm{Y}}_{A}\left\{i\right\}$ are the same. Then, we have
	\begin{equation}\label{q2}
	{{\bm{Y}}_{A'}}\left\{ i \right\}\left[ {1,:} \right] = {{\bm{Y}}_A}\left\{ i \right\}\left[ {1,:} \right].
	\end{equation}
	Similar differential encoding method can be used for the received phase spectrum set ${\bm{Y}}_P$. For the $i^{\rm th}$ matrix in ${\bm{Y}}_P$, i.e., ${{\bm{Y}}_{P}}\left\{ i \right\}$, we obtain the differential encoded matrix ${{\bm{Y}}_{P'}}\left\{ i \right\}$ by
	\begin{equation}\label{q3}
	{{\bm{Y}}_{P'}}\left\{ i \right\}\left[ {j,:} \right] = {{\bm{Y}}_P}\left\{ i \right\}\left[ {j ,:} \right] - {{\bm{Y}}_P}\left\{ i \right\}\left[ {j - 1,:} \right],
	\end{equation}
	where $j = 2,\ldots,L$. Considering that the phase response value of the RIS is added to the signal phase value, we let the first column in ${\bm{Y}}_{P'}\left\{i\right\}$ be the first column in ${\bm{Y}}_{P}\left\{i\right\}$ minus the phase response of the first transmissive element as
	\begin{equation}\label{q4}
	{{\bm{Y}}_{P'}}\left\{ i \right\}\left[ {1,:} \right] = {{\bm{Y}}_P}\left\{ i \right\}\left[ {1,:} \right] - {\bm{\Phi}}_P^{\left(i\right)}\left[ {1,:} \right].
	\end{equation}
	To use the amplitude response matrix of the RIS as the prior knowledge, we multiply the amplitude response matrix of the RIS at the $i^{\rm th}$ moment and ${\bm{Y}}_{P'}\left\{i\right\}$ by elements as
	\begin{equation}\label{q5}
	{{\bm{Y}}_{P'}}\left\{ i \right\} = {{\bm{Y}}_{P'}}\left\{ i \right\} \circ {\bm{\Phi}}_A^{\left(i\right)}.
	\end{equation}
	In addition to the steps of \eqref{q1}, \eqref{q2}, \eqref{q3}, \eqref{q4}, and \eqref{q5}, the transmissive elements on the RIS should be designed by following Remark~\ref{rem1} to make the amplitude response matrix of the RIS available as a special coded aperture, i.e., prior knowledge used in decoding.
	\begin{rem}\label{rem1}
	To achieve differential encoding, we should let every transmissive element on the RIS have the same hardware structure. Thus, different transmissive elements have the same amplitude and phase response to the signals with the same frequency, as shown in Fig.~\ref{Compress} (Part I). Specifically, every column in \eqref{CodeA} and \eqref{CodeP} is the same. This ensures that each column of ${\bm{Y}}_{A'}\left\{i\right\}$ can be represented as the signal amplitude difference values multiplied by the amplitude response values of the RIS as in \eqref{re1a}.
	\end{rem}
	Then, we can express ${\bm{Y}}_{A'}\left\{i\right\}$ and ${\bm{Y}}_{P'}\left\{i\right\}$ as
	\begin{equation}\label{re1a}
	{\bm{Y}}_{A'}\left\{i\right\} = {\bm{H}}_{A'}^{\left(i\right)} \circ {\bf \Phi}_{A}^{\left(i\right)},
	\end{equation}
	and
	\begin{equation}
	{\bm{Y}}_{P'}\left\{i\right\} = {\bm{H}}_{P'}^{\left(i\right)} \circ {\bf \Phi}_{A}^{\left(i\right)},
	\end{equation}
	where ${\bm{H}}_{A'}^{\left(i\right)}$ and ${\bm{H}}_{P'}^{\left(i\right)}$ are the $i^{\rm th}$ differential encoded amplitude and phase spectrums, respectively, and ${\bf \Phi}_{A}^{\left(i\right)}$ can be regarded as the corresponding codebook.

	\begin{algorithm}[t]
	{\small \caption{The algorithm for inverse semantic-aware encoding.} 
	\label{Algorithm1}
	\hspace*{0.02in} {\bf Input:}
	The received amplitude and phase spectrums in the active sensors: ${\bm{Y}}_{A}$ and ${\bm{Y}}_{P}$\\
	\hspace*{0.02in} {\bf Output:}
	The amplitude and phase MetaSpectrums: $\bm{Z}_A$ and $\bm{Z}_P$
	\begin{algorithmic}[1]
	\State {\textit{\#\# Achieve differential encoding}}
	\For{Every ${\bm{Y}}_A\left\{i\right\}$ in ${\bm{Y}}_{A}$}
	\State Obtain ${\bm{Y}}_{A'}\left\{i\right\}$ according to \eqref{q1} and \eqref{q2}
	\EndFor
	\For{Every ${\bm{Y}}_P\left\{i\right\}$ in ${\bm{Y}}_{P}$}
	\State Obtain ${\bm{Y}}_{P'}\left\{i\right\}$ according to \eqref{q3}, \eqref{q4}, and \eqref{q5}
	\EndFor
	\State {\textit{\#\# Achieve shifting addition compression}}
	\State Use ${{\bm{Y}}_{A'}}$ to obtain ${\bm X}_A$ according to \eqref{XA}
	\State Use ${{\bm{Y}}_{P'}}$ to obtain ${\bm X}_P$ according to \eqref{XP}
	\State Obtain amplitude MetaSpectrum $\bm{Z}_A$ according to \eqref{sum}
	\State Obtain phase MetaSpectrum $\bm{Z}_P$ according to \eqref{sumP}
	\State \Return $\bm{Z}_A$ and $\bm{Z}_P$
	\end{algorithmic}}
	\end{algorithm}
	\subsubsection{Shifting Addition}
	To replace the spatial shifting operation to the object spectrum that is performed by a dispersing lens in the SCI system, we perform zero compensation processing to the amplitude and phase spectrums as follows:
	\begin{equation}\label{XA}
	{{\bm{X}}_A} = \!\left\{\! {\left[\!\!\! {\begin{array}{*{20}{c}}
	{{{\bm{Q}}_1}\left( 1 \right)} \\ 
	{{{\bm{Y}}_{A'}}\left\{ 1 \right\}} \\ 
	{{{\bm{Q}}_2}\left( 1 \right)} 
	\end{array}} \!\!\!\right]\!\!, \!\ldots \!,\!\!\left[\!\!\! {\begin{array}{*{20}{c}}
	{{{\bm{Q}}_1}\left( i \right)} \\ 
	{{{\bm{Y}}_{A'}}\left\{ i \right\}} \\ 
	{{{\bm{Q}}_2}\left( i \right)} 
	\end{array}} \!\!\!\right]\!\!, \!\ldots \!,\!\!\left[\!\!\! {\begin{array}{*{20}{c}}
	{{{\bm{Q}}_1}\left( T \right)} \\ 
	{{{\bm{Y}}_{A'}}\left\{ T \right\}} \\ 
	{{{\bm{Q}}_2}\left( T \right)} 
	\end{array}} \!\!\!\right]} \!\right\},
	\end{equation}
	where $ {\bm{Q}_1}\left( i \right) \in {\mathbb{R}^{\left( {i - 1} \right) D \times L}} $, $ {\bm{Q}_2}\left( i \right) \in {\mathbb{R}^{ \left( {T  - i} \right) D \times L}} $, ${{\bm{X}}_A} \in {\mathbb{R}^{\left( D\left( {T - 1} \right) + K\right)  \times L}}$, every elements in both ${\bm{Q}_1}$ and ${\bm{Q}_2}$ is zero, and $D$ is the unit displacement step.
	
	Thus, the amplitude MetaSpectrum, $\bm{Z}_A$, can be obtained by 
	\begin{equation}\label{sum}
	\bm{Z}_A = \sum\limits_{i = 1}^T {{{\bm{X}}_A}\left\{ i \right\}},
	\end{equation}
	where $\bm{Z}_A \in {\mathbb{R}^{\left( K + \left( {T - 1} \right) D\right) \times  L}} $ can be transmitted or stored. Similarly, the phase MetaSpectrum, $\bm{Z}_P$, can be expressed as 
	\begin{equation}\label{sumP}
	\bm{Z}_P = \sum\limits_{i = 1}^T {{{\bm{X}}_P}\left\{ i \right\}},
	\end{equation}
	where
	\begin{equation}\label{XP}
	{{\bm{X}}_P} = \!\left\{\! {\left[\!\!\! {\begin{array}{*{20}{c}}
	{{{\bm{Q}}_1}\left( 1 \right)} \\ 
	{{{\bm{Y}}_{P'}}\left\{ 1 \right\}} \\ 
	{{{\bm{Q}}_2}\left( 1 \right)} 
	\end{array}} \!\!\!\right]\!\!, \!\cdots \!,\!\!\left[\!\!\! {\begin{array}{*{20}{c}}
	{{{\bm{Q}}_1}\left( i \right)} \\ 
	{{{\bm{Y}}_{P'}}\left\{ i \right\}} \\ 
	{{{\bm{Q}}_2}\left( i \right)} 
	\end{array}} \!\!\!\right]\!\!, \!\cdots \!,\!\!\left[\!\!\! {\begin{array}{*{20}{c}}
	{{{\bm{Q}}_1}\left( T \right)} \\ 
	{{{\bm{Y}}_{P'}}\left\{ T \right\}} \\ 
	{{{\bm{Q}}_2}\left( T \right)} 
	\end{array}} \!\!\!\right]} \!\right\}.
	\end{equation}
	
	The overall RIS-aided encoding method is in {\bf{Algorithm~\ref{Algorithm1}}}, which has polynomial complexity. After the RIS-aided encoding, we observe that the sensing data volume is significantly reduced. {\color{black}
To demonstrate the efficiency of the inverse semantic communications approach, we define the data compression ratio, i.e., $\rho $, as the ratio between the number of elements in the received amplitude and phase spectrums and the number of elements in the coded MetaSpectrums. We provide the analysis of $\rho$ in {\bf{Proposition~\ref{P1}}}.
\begin{prop}\label{P1}
	The data compression ratio $\rho $ of our proposed inverse semantic-aware coding method is approximately ${1}/{T}$.
	\begin{IEEEproof}
		The total number of elements in the $T$ recorded signal amplitude and phase spectrum is $2KLT$. Meanwhile, the total number of elements in $\bm{Z}_A$ and $\bm{Z}_P$ is $2{\left( {K + \left( {T - 1} \right)D} \right) \times L}$. As a result, $\rho$ can be calculated as:
		\begin{equation}\label{tario}
			\rho  = \frac{2{\left( {K + \left( {T - 1} \right)D} \right) \times L}}{2{KLT}}{\text{ = }}\frac{1}{T} + \left( {1 - \frac{1}{T}} \right)\frac{D}{K}.
		\end{equation}
		Given that $D$ is generally small compared to $K$, for instance, $D = 1$ in \cite{meng2021self} and $K=2048$ in the IEEE 802.11ax protocol, the term $\left( {1 - \frac{1}{T}} \right)\frac{D}{K}$ in~\eqref{tario} can be negligible. Consequently, the value of $\rho$ is approximately ${1}/{T}$, which completes the proof.
	\end{IEEEproof}
\end{prop}	
}
	
	Note that in the above discussion, we encode $T$ amplitude or phase spectrums into one spectrum. However, the $T$ spectrums does not need and should not be sensed continuously in time. The reason is that  the wireless channel remains stable during the channel coherence time. Specifically, as the moving or action speed of people is limited, the CSI within the channel coherence time can be considered as constant without loss of precision~\cite{vasisht2016decimeter}. Considering that the available maximal sensing frequency of the active sensors is much higher than the required frequency, we next propose a sampling scheme that selects the most relevant spectrum for the completion of sensing tasks over the channel coherence time for recording and encoding.

	\subsection{Semantic Hash Sampling}
	\begin{figure}[t]
	\centering
	\includegraphics[width=0.44\textwidth]{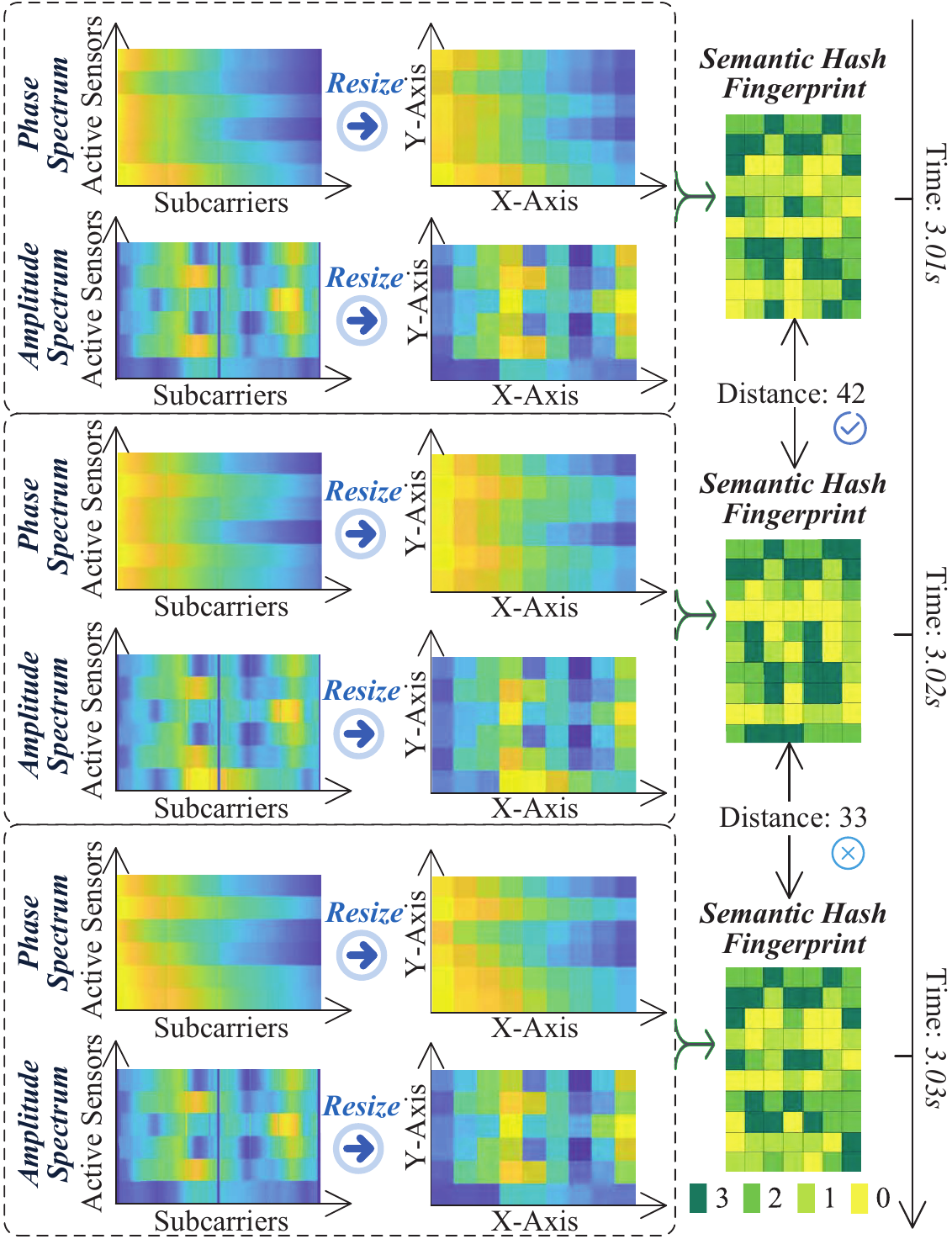} 
	\caption{The process of generating the resized matrices from the amplitude and phase spectrums, and then obtaining the semantic hash fingerprint.}
	\label{hash}
	\end{figure}
We divide the time into segments. Without loss of generality, we consider that the active sensors can perform $T_N$ times sensing in one time segment. From each time segment, one pair of amplitude and phase spectrums is selected to record. In the $i^{\rm th}$ time segment, we express $T_N$ received amplitude and $T_N$ phase spectrums as two sets, i.e., {\small ${\bm{S}}_{A_i} = \left\{ {\bm{H}}_{A_i}^{\left(k\right)} \circ {\bm{\Phi}} _{A_i}^{\left(k\right)}\right\}$} $\left( k = 1,\ldots,T_N\right) $ and {\small ${\bm{S}}_{P_i} = \left\{ {\bm{H}}_{P_i}^{\left(k\right)} \circ {\bm{\Phi}} _{P_i}^{\left(k\right)}\right\}$}, respectively. The recorded amplitude and phase spectrums that are selected from the $i^{\rm th}$ time segment are ${\bm{Y}}_{A}{\left\{i\right\}}$ $\left( i = 1,\ldots,T\right) $ and ${\bm{Y}}_{P}{\left\{i\right\}}$, respectively.
	
To remove information that is not relevant to the task, the traditional method is uniform sampling that selects the first pair of amplitude and phase spectrums in each time segment to record. However, we cannot guarantee that the first pair in every segment is always the most informative pair. Therefore, a better solution is to use one indicator to judge the semantic information richness of the pair of spectrums. As we discussed in Section~\ref{CM}, the information related to sensing tasks is contained in the changes of amplitude and phase spectrums. Therefore, we can select the pair of spectrums that has the largest change compared to the previous signal spectrums in each time segment. Note that the mean square error (MSE) is not recommended to be used as the indicator to compare the difference between spectrums. The reasons are given as follows:
	\begin{itemize}
	\item The MSE is calculated using the absolute values of the signal amplitude. However, the absolute values are not important for sensing tasks. The critical information is in the changing process of the signal amplitude over time~\cite{niu2022rethinking}.
	\item The results of MSE may be affected by several outliers, i.e., signal amplitude fluctuation at a certain time caused by the interference.
	\item Because the number of elements in the signal spectrum is large, calculating MSE brings large resource consumption.
	\end{itemize}
	Therefore, we have to propose a new indicator to characterize the semantic information richness in signal spectrums. Considering the success of the perceptual image hashing method~\cite{du2020perceptual} in the field of image retrieval, we aim to use a string of characters, i.e., fingerprints, to characterize the amplitude and phase spectrums. Perceptual image hashing~\cite{du2020perceptual} is a family of algorithms that generate content-based image hash fingerprints. Then, the Hamming distance between two fingerprints can be used to quantify the similarity of two images. The larger the Hamming distance is, the smaller the similarity of the images have. Although the hash fingerprints can be calculated efficiently with low energy cost, it cannot be applied directly to the similarity detection of sensing data. The reason is that, at each moment, we have one amplitude spectrum and one phase spectrum as shown in Fig.~\ref{hash}, which are both required to achieve sensing tasks. Thus, we propose a novel four-level semantic hash sampling method in {\bf{Algorithm~\ref{Algorithm2}}} to select task-related signals spectrums for encoding, which is used before {\bf{Algorithm~\ref{Algorithm1}}}.
	
	As shown in {\bf{Algorithm~\ref{Algorithm2}}}, to obtain the semantic hash matrices, the first step is to resize the $T_K$ amplitude and phase spectrums. The purpose is to produce a small data size, which hastens the processing time~\cite{tuncer2020novel} and preserves the features of the spectrums. Similar to the image pHash method~\cite{khanam2018implementation}, we calculate the average values of the resized amplitude and phase matrices. Different from the conventional hash method, we define four values, i.e., 0, 1, 2, and 3, as values in the hash fingerprints. Thus, we perform the operations as shown in lines $8 - 15$ of {\bf{Algorithm~\ref{Algorithm2}}} to convert the spectrums to semantic hash fingerprints in polynomial complexity. For the $k^{\rm th}$ pair of amplitude and phase spectrums, we use the Hamming distance, which measures the number of different values, between the $k^{\rm th}$ hash fingerprint and the $\left( k-1\right) ^{\rm th}$ one to indicate the semantic information richness of the $k^{\rm th}$ spectrums. Therefore, we can record the pair of spectrums that have the largest Hamming distance to the previous pair of spectrums. 
	{\color{black}
The complexity analysis for each step in our proposed algorithm is as follows:
\begin{itemize}
	\item {\textit{Obtaining the semantic hash matrix set:}} The resizing operation~\cite{khanam2018implementation} has a complexity of $\mathcal{O}(R_x R_y)$ per matrix. Therefore, the complexity of obtaining the semantic hash matrix set for $T_K$ spectrums is $\mathcal{O}(T_K R_x R_y)$.
	\item {\textit{Calculating the Hamming distance:}} The complexity consists of $O(1)$ for creating the vector $\bm{\mathcal{D}}$, $O(T_K)$ for the outer loop iterating through ${\bm{\mathcal{H}}}_{i}$, and $O(R_x R_y)$ for the inner loop iterating through ${\bm{\mathcal{H}}}_{i}{\left\{k\right\}}$. This results in a total complexity of $\mathcal{O}(T_K R_x R_y)$.
	\item {\textit{Selecting the spectrum and recording the information richness:}} The complexity is dominated by the linear search to find $k_{\max}$, which maximizes $\bm{\mathcal{D}}(k)$. As the vector is unsorted, the complexity is $\mathcal{O}(T_K)$.
\end{itemize}
These steps are performed sequentially, leading to the total complexity of $\mathcal{O}\left( T_K \left( 2 R_x R_y + 1\right) \right)$. Since the constant terms do not significantly affect the growth rate, the total complexity of {\bf{Algorithm~\ref{Algorithm2}}} simplifies to $\mathcal{O}(T_K R_x R_y)$.	
}

	\begin{algorithm}[t]
	{\small \caption{The algorithm for semantic hash sampling}
	\label{Algorithm2}
	\hspace*{0.02in} {\bf Input:}
	\begin{itemize}
	\item The received amplitude and phase spectrums sets in the $i^{\rm th}$ time segment: ${\bm{S}}_{A_i}$ and ${\bm{S}}_{P_i}$
	\item The dimensions of the resized matrices: $R_x$ and $R_y$
	\end{itemize}
	\hspace*{0.02in} {\bf Output:}
	The selected amplitude and phase spectrums: ${\bm{Y}}_{A}{\left\{i\right\}}$ and ${\bm{Y}}_{P}{\left\{i\right\}}$
	\begin{algorithmic}[1]
	\State {\textit{\#\# Obtain the semantic hash matrix set}} 
	\State Create an empty matrix set ${\bm{\mathcal{H}}}_{i} \in {\mathbb{R}}^{R_x\times{R_y}\times T_K} $ to record the semantic hash values
	\For{Every {\small ${\bm{S}}_{A_i}{\left\{k\right\}}$} in {\small ${\bm{S}}_{A_i}$}}
	\State Obtain amplitude and phase spectrums ${{\bm{H}}}_{A_i}^{\left(k\right)}$ and ${\bm{H}}_{P_i}^{\left(k\right)}$ with the prior knowledge ${\bm{\Phi}} _{A_i}^{\left(k\right)}$ and ${\bm{\Phi}} _{P_i}^{\left(k\right)}$, respectively
	\State Resize ${\bm{H}}_{A_i}^{\left(k\right)}$ and ${\bm{H}}_{P_i}^{\left(k\right)}$ into small matrices ${\bm{h}}_{A_i}^{\left(k\right)}\in {\mathbb{R}}^{R_x\times{R_y}}$ and ${\bm{h}}_{P_i}^{\left(k\right)}\in {\mathbb{R}}^{R_x\times{R_y}}$, respectively
	\State Calculate the average values of ${\bm{h}}_{A_i}^{\left(k\right)}$ and ${\bm{h}}_{P_i}^{\left(k\right)}$, denoted as ${{h}}_{A_i}^{\left(k\right)}$ and ${{h}}_{P_i}^{\left(k\right)}$, respectively
	\For{Every element pair in ${\bm{h}}_{A_i}^{\left(k\right)}$ and ${\bm{h}}_{P_i}^{\left(k\right)}$}
	\If{${{\bm{h}}}_{A_i}^{\left(k\right)}\left[ {x,y} \right] \geqslant {{h}}_{A_i}^{\left(k\right)}$ and  ${{\bm{h}}}_{P_i}^{\left(k\right)}\left[ {x,y} \right] \geqslant {{h}}_{P_i}^{\left(k\right)}$}
	\State Let ${\bm{\mathcal{H}}}_{i}{\left\{k\right\}}\left[ {x,y} \right] \leftarrow 3$
	\ElsIf{${{\bm{h}}}_{A_i}^{\left(k\right)}\left[ {x,y} \right] \geqslant {{h}}_{A_i}^{\left(k\right)}$ and ${{\bm{h}}}_{P_i}^{\left(k\right)}\left[ {x,y} \right] < {{h}}_{P_i}^{\left(k\right)}$}
	\State Let ${\bm{\mathcal{H}}}_{i}{\left\{k\right\}}\left[ {x,y} \right] \leftarrow 2$
	\ElsIf{${{\bm{h}}}_{A_i}^{\left(k\right)}\left[ {x,y} \right] < {{h}}_{A_i}^{\left(k\right)}$ and ${{\bm{h}}}_{P_i}^{\left(k\right)}\left[ {x,y} \right] \geqslant {{h}}_{P_i}^{\left(k\right)}$}
	\State Let ${\bm{\mathcal{H}}}_{i}{\left\{k\right\}}\left[ {x,y} \right] \leftarrow 1$
	\Else 
	\State Let ${\bm{\mathcal{H}}}_{i}{\left\{k\right\}}\left[ {x,y} \right] \leftarrow 0$
	\EndIf
	\EndFor
	\EndFor
	\vspace{0.01cm}
	\State {\textit{\#\# Calculate the Hamming distance}}
	\State Create an empty vector ${\bm{\mathcal{D}}} \in {\mathbb{R}}^{1\times{T_K}}$ to record the Hamming distance values
	\For{Every ${\bm{\mathcal{H}}}_{i}{\left\{k\right\}}$ in ${\bm{\mathcal{H}}}_{i}$}
	\State Create a temporary variable $d$
	\For{Every element in ${\bm{\mathcal{H}}}_{i}{\left\{k\right\}}$}
	\If{The element value is different from the element value in the same position in ${\bm{\mathcal{H}}}_{i}{\left\{k-1\right\}}$}
	\State $d \leftarrow d+1$
	\EndIf
	\EndFor
	\State ${\bm{\mathcal{D}}} \left( k \right)  \leftarrow d $.
	\EndFor
	\vspace{0.01cm}
	\State {\textit{\#\# Select the spectrum and record the information richness}}
	\State Find $k_{\max}$ that maximizes ${\bm{\mathcal{D}}} \left( k \right)$, i.e., ${\bm{\mathcal{D}}} \left( k_{\max} \right) = \max\left\{ {\bm{\mathcal{D}}} \right\}$
	\State Record the value of ${\bm{\mathcal{D}}} \left( k_{\max} \right)$
	\State Let $i = k_{\max}$
	\State \Return ${\bm{Y}}_{A}{\left\{i\right\}}$ and ${\bm{Y}}_{P}{\left\{i\right\}}$
	\end{algorithmic}}
	\end{algorithm}
	
	\section{Inverse Semantic-aware Decoding}\label{SS5}
	In this section, we propose the inverse semantic-aware self-supervised decoding method. We also introduce how to use the recovered signal spectrums for 2D AoA and ToF estimation, which supports various sensing tasks.
	
	\subsection{Objective Function}
	We first rewrite the amplitude and phase MetaSpectrums $\bm{Z}_A$ and $\bm{Z}_P$ in the vectorized formulations, respectively. Let $\rm{vec}\left(\cdot\right) $ denote the matrix vectorization operation that concatenates columns into one vector, and $\rm{diag}\left({\bf{a}}\right) $ denote the operation of converting the vector $\bf{a}$ into a diagonal matrix where the diagonal element is $\bf{a}$. As such, we rewrite the matrix formulations \eqref{sum} and \eqref{sumP} as
	\begin{equation}\label{vec}
	{\bf z}_A = {\bf \Phi} {\bf x}_A,
	\end{equation}
	and
	\begin{equation}\label{vecP}
	{\bf z}_P = {\bf \Phi} {\bf x}_P,
	\end{equation}
	respectively, where $\bm{z}_A = {\rm{vec}}\left({\bm{Z}}_A\right)$, ${\bm{z}}_P = {\rm{vec}}\left({\bm{Z}}_P\right)$, $\bm{z}_A$ and $\bm{z}_P \in {\mathbb{R}^{\left( K + \left( {T - 1} \right) D\right) L \times 1}} $,  $\bm{x}_A = \left[\bm{x}_{1,A}^{\rm T} \cdots {\bm{x}}_{i,A}^{\rm T} \cdots \bm{x}_{T,A}^{\rm T}\right]^{\rm T}$, $\bm{x}_P = \left[\bm{x}_{1,P}^{\rm T} \cdots {\bm{x}}_{i,P}^{\rm T} \cdots \bm{x}_{T,P}^{\rm T}\right]^{\rm T}$, $\bm{x}_{i,A} = {\rm{vec}}\left({{{\bm{H}}_{A'}^{\left( i\right) }}} \right)$, $\bm{x}_{i,P} = {\rm{vec}}\left({{{\bm{H}}_{P'}^{\left( i\right) }}} \right)$, ${\bm{x}}_A$ and ${\bm{x}}_P \in {\mathbb{R}^{TKL\times 1}} $,
	\begin{equation}\label{xianyan}
	{\bf \Phi} = \left[ {\begin{array}{*{20}{c}}
	{{{\bm{Q}}_3}\left( 1 \right)}& \cdots &{{{\bm{Q}}_3}\left( i \right)}& \cdots &{{{\bm{Q}}_3}\left( T \right)} \\ 
	{{\bm \phi} _A^{\left( 1 \right)}}& \ddots &{{\bm \phi} _A^{\left( i \right)}}& \ddots &{{\bm \phi} _A^{\left( T \right)}} \\ 
	{{{\bm{Q}}_4}\left( 1 \right)}& \cdots &{{{\bm{Q}}_4}\left( i \right)}& \cdots &{{{\bm{Q}}_4}\left( T \right)} 
	\end{array}} \right],
	\end{equation}
	${{{\bm{Q}}_3}\left( 1 \right)} \in {\mathbb{R}^{\left(i-1\right) DL\times KL}}$, ${{{\bm{Q}}_4}\left( 1 \right)} \in {\mathbb{R}^{\left(T-1\right) DL \times KL}}$, every element in both ${\bm{Q}_3}$ and ${\bm{Q}_4}$ is zero, ${\bm \phi}_A^{\left( i \right)} = {\rm diag}\left( {{\rm vec}\left( {{\bf \Phi}_A^{\left( i \right)}} \right)} \right) $, ${\bm \phi}_A^{\left( i \right)} \in {\mathbb{R}^{KL \times KL}}$, and ${\bf \Phi} \in {\mathbb{R}^{{\left( K + \left( {T - 1} \right) D\right) L} \times TKL}}$.
	
	Note that ${\bf \Phi}$ can be obtained using \eqref{xianyan} and the prior knowledge, i.e., the amplitude response matrices of the RIS, and ${\bf z}_A$ and ${\bf z}_P$ are the known vectorized amplitude and phase MetaSpectrums, respectively. Our goal is to decode ${\bf x}_A$ and ${\bf x}_P$ from ${\bf z}_A$ and ${\bf z}_P$, respectively. Although this problem is related to CS, most theories that are developed for CS cannot be used because that the matrix ${\bf \Phi}$ follows a very specific structure as~\eqref{xianyan}. Fortunately, solid theoretical proof has shown that both ${\bf x}_A$ in \eqref{vec} and ${\bf x}_P$ in \eqref{vecP} can be recovered even when $T>1$ \cite{jalali2019snapshot}.
	
	The decoding objective function can be formulated as
	\begin{equation}\label{utility}
	\mathop {\min }\limits_{{{\bm{x}}_A},{{\bm{x}}_P}}\! \alpha_1 \!\left\| {{{\bm{z}}_A} \!-\! {{\bm{\Phi }}}{{\bm{x}}_A}} \right\|^2 + \!\alpha_2 \!\left\| {{{\bm{z}}_P}\! - \!{{\bm{\Phi }}}{{\bm{x}}_P}} \right\|^2\!,
	\end{equation}
	where $\alpha_1$ and $\alpha_2$ are the balance parameters that can be selected according to the specific wireless sensing task. For example, heartbeat and breath detection requires higher accuracy for the phase spectrum~\cite{liu2019wireless}, and the amplitude spectrum is more significant in sensing tasks such as intrusion or fall detection~\cite{ramadan2020efficient}. We propose the algorithm for solving \eqref{utility} as follows.
	
	\subsection{Self-supervised Decoding Method}
	To solve~\eqref{utility}, although different hand-crafted priors, e.g., total variation and sparsity, can be added as the regularization term to improve the decoding performance, it is hard to choose a suitable prior that fits the differential encoded amplitude and phase spectrums ${{\bm{x}}_A}$ and ${{\bm{x}}_P}$. Motivated by the success of deep convolutional neural networks (ConvNets) in inverse problems such as single-image super-resolution~\cite{ledig2017photo} and denoising~\cite{lefkimmiatis2017non}, we use the implicit prior captured by the ConvNets, e.g., deep image prior~\cite{ulyanov2018deep,mataev2019deepred}, to achieve self-supervised decoding\footnote{Another solution is to design a suitable explicit regularization term for decoding sensing signal spectrums and use the explicit and implicit priors jointly~\cite{mataev2019deepred}. This is left for the future work.}.
	
	By considering that the unknown amplitude and phase spectrums are the outputs of neural networks, i.e., ${{\bm{\mathcal T}}\!_{{{\bm{\Theta }}_A}}}\!\!\left( {\bm{e}} \right)$ and ${{\bm{\mathcal T}}\!_{{{\bm{\Theta }}_P}}}\!\!\left( {\bm{e}} \right)$, respectively, the decoding problem~\eqref{utility} can be re-written as
	\begin{equation}\label{utility2}
	\begin{array}{*{20}{l}}
	{\mathop {\min }\limits_{{{\bm{\Theta}} _A},{{\bm{\Theta}} _P}}}&\!\!\!\!{{\alpha _1}\!\left\| {{{\bm{z}}_A} \!-\! {{\bm{\Phi }}}{{\bm{\mathcal T}}\!_{{{\bm{\Theta }}_A}}}\!\!\left( {\bm{e}} \right)} \right\|^2 \!+ \!{\alpha _2}\!\left\| {{{\bm{z}}_P}\! - \!{{\bm{\Phi }}}{{\bm{\mathcal T}}\!_{{{\bm{\Theta }}_P}}}\!\!\left( {\bm{e}} \right)} \right\|^2,} \\ 
	{\:\:\:{\rm{s.t.}}}&\!\!\!\!{{{{\bm{\hat x}}}_A} = {{\bm{\mathcal T}}\!_{{{\bm{\Theta }}_A}}}\!\!\left( {\bm{e}} \right)}, \\ 
	{}&\!\!\!\!{{{{\bm{\hat x}}}_P} = {{\bm{\mathcal T}}\!_{{{\bm{\Theta }}_P}}}\!\!\left( {\bm{e}} \right)},
	\end{array}
	\end{equation}
	where ${{\bm{\Theta }}_A}$ and ${{\bm{\Theta }}_P}$ are the parameters of networks to be learned, and ${\bm{e}}$ is a random vector. Since the training of ${{\bm{\Theta }}_A}$ and ${{\bm{\Theta }}_P}$ is part of the decoding process, this procedure is self-supervised and no pre-training process is required.

	To solve the problem~\eqref{utility2}, we introduce two auxiliary variables ${\bm{t}_1}$ and ${\bm{t}_2}\in {\mathbb R}^{TKL}$, and corresponding weight parameters $\beta_1$ and $\beta_2$. Then, the constraints can be turned into penalty terms using the augmented Lagrangian method~\cite{afonso2010fast} as 
	\begin{equation}\label{finalpro}
	\begin{array}{*{20}{l}}
	{\mathop {\min }\limits_{{\bm{\Theta}}_A,{\bm{\Theta}}_P,{\bm{x}}_A,{\bm{x}}_P}}
	&{\bm{\mathcal{F}}}_1\left({\bm{\Theta}}_A, {\bm{x}}_A\right) + {\bm{\mathcal{F}}}_2\left({\bm{\Theta}}_P, {\bm{x}}_P\right),
	\end{array}
	\end{equation}
	where
	\begin{equation}
	{\bm{\mathcal{F}}}_1\!\! =\! {\alpha _1}\!\!\left(\! \left\| {{{\bm{z}}_A} \!-\! {{\bm{\Phi }}}\!{{\bm{\mathcal T}}\!_{{{\bm{\Theta }}_A}}}\!\!\left( {\bm{e}} \right)} \right\|^2 \!+\!{\beta _1}\!\left\| {{{\bm{x}}_A} \! - \! {{\bm{\mathcal T}}\!_{{{\bm{\Theta }}_A}}}\!\!\left( {\bm{e}} \right) \!- \!{\bm{t}}_1} \right\|^2\!\right)\!,
	\end{equation}
	and
	\begin{equation}
	{\bm{\mathcal{F}}}_2\!\!=\! {\alpha _2}\!\!\left(\! \left\| {{{\bm{z}}_P} \!-\! {{\bm{\Phi }}}{{\bm{\mathcal T}}\!_{{{\bm{\Theta }}_P}}}\!\!\left( {\bm{e}} \right)} \right\|^2 \!+\!{\beta _2}\!\left\| {{{\bm{x}}_P} \! - \! {{\bm{\mathcal T}}\!_{{{\bm{\Theta }}_P}}}\!\!\left( {\bm{e}} \right) \!- \!{\bm{t}}_2} \right\|^2\!\right)\!.
	\end{equation}
	
	With the help of the alternating direction method of multipliers (ADMM)~\cite{boyd2011distributed}, the problem~\eqref{finalpro} can be solved by a sequential update of the six variables, i.e., ${\bm{\Theta}}_A$, ${\bm{\Theta}}_P$, ${\bm{x}}_A$, ${\bm{x}}_P$, ${\bm{t}}_1$, and ${\bm{t}}_2$.
	
	{\it{1) The update of ${\bm{\Theta}}_A$ while fixing other variables:}}
	\begin{equation}\label{s1}
	\begin{array}{*{20}{l}}
	{\mathop {\min }\limits_{{\bm{\Theta}}_A}}
	&\!\!\! \left\| {{{\bm{z}}_A} \!-\! {{\bm{\Phi }}}{{\bm{\mathcal T}}\!_{{{\bm{\Theta }}_A}}}\!\!\left( {\bm{e}} \right)} \right\|^2 \!+\!{\beta _1}\!\left\| {{{\bm{x}}_A} \! - \! {{\bm{\mathcal T}}\!_{{{\bm{\Theta }}_A}}}\!\!\left( {\bm{e}} \right) \!- \!{\bm{t}}_1} \right\|^2\!,
	\end{array}
	\end{equation}
	which can be solved using the steepest descent and back-propagation optimization methods~\cite{ulyanov2018deep}. Note that ${\beta _1}\!\left\| {{{\bm{x}}_A} \! - \! {{\bm{\mathcal T}}\!_{{{\bm{\Theta }}_A}}}\!\!\left( {\bm{e}} \right) \!- \!{\bm{t}}_1} \right\|^2$ in~\eqref{s1} can be regarded as the denoising of ${{\bm{x}}_A} \! - \! {{\bm{t}}_1}$, which also serves as a proximity regularization that forces ${{\bm{\mathcal T}}\!_{{{\bm{\Theta }}_A}}}\!\!\left( {\bm{e}} \right)$ to be close to ${{\bm{x}}_A} \! - \! {{\bm{t}}_1}$. This second term provides additional stabilizing and robustifying effect to the back-propagation method.
	
	{\it{2) The update of ${\bm{x}}_A$ while fixing other variables:}}
	\begin{equation}\label{Q2}
	\begin{array}{*{20}{l}}
	{\mathop {\min }\limits_{{\bm{x}}_A}}
	&\!\!\!\left\| {{{\bm{x}}_A} \! - \! {{\bm{\mathcal T}}\!_{{{\bm{\Theta }}_A}}}\!\!\left( {\bm{e}} \right) \!- \!{\bm{t}}_1} \right\|^2\!,
	\end{array}
	\end{equation}
	which can be regarded as a denoising problem for ${{\bm{\mathcal T}}\!_{{{\bm{\Theta }}_A}}}\!\!\left( {\bm{e}} \right) + {\bm{t}}_1$. Thus, we have 
	\begin{equation}\label{S2}
	{\bm{\hat x}}_A ={\bm {\mathcal D}}\left({{{\bm{\mathcal T}}_{{{\bm{\Theta}}_A }}}}\!\!\left( {\bm{e}} \right) + {\bm{t}}_1\right),
	\end{equation}
	where ${\bm {\mathcal D}}\left(\cdot\right) $ represents the denoising operator that could be well-studied plug-and-play algorithms~\cite{sreehari2016plug} or a simpler steepest-descent (SD) operator. We present the update equation for SD method as
	\begin{equation}\label{S2P}
	{\bm{ x}}_A^{\left(j + 1\right) } ={\bm{ x}}_A^{\left(j \right) } - s \left({\bm{ x}}_A^{\left(j\right) } - {{{\bm{\mathcal T}}_{{{\bm{\Theta}}_A }}}}\!\!\left( {\bm{e}} \right) - {\mathbf{t}}_1\right),
	\end{equation}
	where $s$ is the steepest-descent step size, and $j$ is the inner loop iteration number.
	
	
	{\it{3) The update of ${\bm{t}}_1$ while fixing other variables:}}
	Because ${\bm{t}}_1$ can be regarded as the Lagrange multipliers vector, ${\bm{t}}_1$ can be updated according to the augmented Lagrangian method~\cite{afonso2010fast} as
	\begin{equation}\label{S3}
	{\bm{t}}_1^{\left( k+1\right) } = {\bm{t}}_1^{\left( k\right) } + {{\bm{\mathcal T}}_{{{\bm{\Theta}}_A^{\left( k\right) }}}}\!\!\left( {\bm{e}} \right) - {\mathbf{x}}_A^{\left( k\right) },
	\end{equation}
	where $k$ denotes the outer loop iteration number.
	
	{\it{4) The update of ${\bm{\Theta}}_P$ while fixing other variables:}}
	Because the network with parameter ${\bm{\Theta}}_P$ is trained independently, we can update ${\bm{\Theta}}_P$ by solving
	\begin{equation}\label{S4}
	\begin{array}{*{20}{l}}
	{\mathop {\min }\limits_{{\bm{\Theta}}_P}}
	&\!\!\! \left\| {{{\mathbf{z}}_P} \!-\! {{\mathbf{\Phi }}}{{\bm{\mathcal T}}\!_{{{\bm{\Theta }}_P}}}\!\!\left( {\bm{e}} \right)} \right\|^2 \!+\!{\beta _2}\!\left\| {{{\mathbf{x}}_P} \! - \! {{\bm{\mathcal T}}\!_{{{\bm{\Theta }}_P}}}\!\!\left( {\bm{e}} \right) \!- \!{\bm{t}}_2} \right\|^2\!,
	\end{array}
	\end{equation}
	with the same method as in~\eqref{s1}.
	
	{\it{5) The update of ${\bm{x}}_P$ while fixing other variables:}}
	To minimize the difference between ${\mathbf{x}}_P$ and ${{\bm{\mathcal T}}_{{{\bm{\Theta}}_A}}}\!\!\left( {\bm{e}} \right) + {\mathbf{t}}_1$, we can update ${\mathbf{x}}_P$ as
	\begin{equation}\label{S5}
	{\mathbf{\hat x}}_P ={\bm {\mathcal D}}\left({{{\bm{\mathcal T}}_{{{\bm{\Theta}}_P}}}}\!\!\left( {\bm{e}} \right) + {\mathbf{t}}_2\right).
	\end{equation}
	where ${\bm {\mathcal D}}$ is the same kind of denoising operator as~\eqref{S2}.
	
	{\it{6) The update of ${\bm{t}}_2$ while fixing other variables:}}
	According to the augmented Lagrangian method~\cite{afonso2010fast}, ${\bm{t}}_2$ can be updated as
	\begin{equation}\label{S6}
	{\bm{t}}_2^{\left( k+1\right) } = {\bm{t}}_2^{\left( k\right) } + {{\bm{\mathcal T}}_{{{\bm{\Theta}}_P^{\left( k\right) }}}}\!\!\left( {\bm{e}} \right) - {\mathbf{x}}_P^{\left( k\right) }.
	\end{equation}
	
	\begin{algorithm}[t]
	{\small \caption{The algorithm for inverse semantic-aware decoding}
	\label{Algorithm3}
	\hspace*{0.02in} {\bf Input:}
	\begin{itemize}
	\item The weight parameters: $\beta_1$ and $\beta_2$
	\item The number of inner iterations of the denoising operator for updating ${\bm{x}}_A$ and ${\bm{x}}_P$: $N_J$
	\item The steepest-descent parameters for updating ${{\bm{\Theta }}_A}$ and ${{\bm{\Theta }}_P}$, respectively
	\end{itemize}
	\hspace*{0.02in} {\bf Output:}
	The original amplitude and phase spectrums, i.e., ${\mathbf{H}}_{A}^{\left( i\right) }$ and ${\mathbf{H}}_{P}^{\left( i\right) }$ $\left( i = 1,\ldots,T\right) $
	\begin{algorithmic}[1]
	\State {\textit{\#\# Reconstruction of the ${\bm{x}}_A$ and ${\bm{x}}_P$}}
	\State Initialize the iteration number $k=0$
	\State Set ${{\bm{\Theta }}_A}$ and ${{\bm{\Theta }}_P}$ randomly
	\While{Not converged}
	\State Update ${\bm{\Theta}}_A$ by solving~\eqref{s1} using steepest descent and back-propagation methods
	\State Update ${\bm{x}}_A$ according to~\eqref{S2}
	\State Update ${\bm{t}}_1$ according to~\eqref{S3}
	\State Update ${\bm{\Theta}}_P$ by solving~\eqref{S4} using steepest descent and back-propagation methods
	\State Update ${\bm{x}}_P$ according to~\eqref{S5}
	\State Update ${\bm{t}}_2$ according to~\eqref{S6}
	\State Let $k \leftarrow k+1$
	\EndWhile
	\State Record ${\bm{x}}_A$ and ${\bm{x}}_P$ after converged
	\vspace{0.02cm}
	\State {\textit{\#\# Differential decoding}}
	\State Recover ${\mathbf{H}}_{A'}{\left\{i\right\}}$ and ${\mathbf{H}}_{P'}{\left\{i\right\}}$ $\left( i = 1,\ldots,T\right) $ according to the definition of ${\bm{x}}_A$ and ${\bm{x}}_P$, i.e.,~\eqref{vec} and \eqref{vecP}
	\For{Every ${\mathbf{H}}_{A'}{\left\{i\right\}}$ and ${\mathbf{H}}_{P'}{\left\{i\right\}}$}
	\State Create empty ${\bm{H}}_{A}\left\{i\right\}$ and ${\bm{H}}_{P}\left\{i\right\}$ to record the decoded results
	\State Obtain ${\bm{H}}_{A}\left\{i\right\}$ and ${\bm{H}}_{P}\left\{i\right\}$ according to \eqref{req1}, \eqref{req2}, \eqref{req3}, and \eqref{req4}
	\EndFor
	\State \Return ${\mathbf{H}}_{A}{\left\{i\right\}}$ and ${\mathbf{H}}_{P}{\left\{i\right\}}$ $\left( i = 1,\ldots,T\right) $
	\end{algorithmic}}
	\end{algorithm}
	{\bf{Algorithm~\ref{Algorithm3}}} summarizes the steps to perform the aforementioned decoding methods, and then recover the original amplitude and phase spectrums. Specifically, after decoding, we obtain the estimated ${\bm{H}}_{A'}\left\{i\right\}$ and ${\bm{H}}_{P'}\left\{i\right\}$. Let the first column in ${\bm{H}}_{A}\left\{i\right\}$ and ${\bm{H}}_{P}\left\{i\right\}$ be the same as that of ${\bm{H}}_{A'}\left\{i\right\}$ and ${\bm{H}}_{P'}\left\{i\right\}$ as
	\begin{equation}\label{req1}
	{{\bm{H}}_{A}}\left\{ i \right\}\left[ {1,:} \right] = {{\bm{H}}_A'}\left\{ i \right\}\left[ {1,:} \right],
	\end{equation}
	and
	\begin{equation}\label{req2}
	{{\bm{H}}_{P}}\left\{ i \right\}\left[ {1,:} \right] = {{\bm{H}}_{P'}}\left\{ i \right\}\left[ {1,:} \right].
	\end{equation}
	For the second to the last columns ($j = 2,\ldots,L$), we have
	\begin{equation}\label{req3}
	{{\bm{H}}_{A}}\left\{ i \right\}\left[ {j,:} \right] = {{\bm{H}}_{A'}}\left\{ i \right\}\left[ {j - 1,:} \right] + {{\bm{H}}_{A'}}\left\{ i \right\}\left[ {j,:} \right],
	\end{equation}
	and
	\begin{equation}\label{req4}
	{{\bm{H}}_{P}}\left\{ i \right\}\left[ {j,:} \right] = {{\bm{H}}_{P'}}\left\{ i \right\}\left[ {j - 1,:} \right] + {{\bm{H}}_{P'}}\left\{ i \right\}\left[ {j,:} \right].
	\end{equation}
	Note that because of the independent iterative training of the two networks and the use of the ADMM method, $\alpha_1$ and $\alpha_2$ have no effect on the objective function. The running time is mainly taken in updating ${{\bm{\Theta }}_A}$ and ${{\bm{\Theta }}_P}$ since the inner denoising operators work efficiently. In Section~\ref{SS6}, we set the inner iteration numbers of the denoising operators for updating ${\bm{x}}_A$ and ${\bm{x}}_P$ to be both $600$, and the outer loop maximal iteration number is $18$, i.e., $18$ ADMM iterations. The average running time for decoding one MetaSpectrum, which is obtained by encoding $20$ original amplitude spectrums, is about $1$ minute with the experiment setting in Section~\ref{SS6}. {\color{black}While the self-supervised decoding approach may not be optimal for high real-time decoding of sensing signal data, our proposed method can be effectively applied to sensing tasks that require large amounts of historical data storage for analysis, such as healthcare monitoring, sleeping position detection, and historical intrusion or walking behavior analysis.}
	
	With the decoded ${\mathbf{H}}_{A}{\left\{i\right\}}$ and ${\mathbf{H}}_{P}{\left\{i\right\}}$ $\left( i = 1,\ldots,T\right)$, the original signal at each moment can be recovered to the form of complex matrices. Then, the 2D AoA and ToF can be jointly estimated~\cite{hua1989shaped}, which can be used to complete a series of sensing tasks. Thus, the steering matrices of $L$-shaped array in the $x$ and $y$ directions, which describe how the sensor array uses each individual element to select a spatial path for the transmission, can be expressed as
	\begin{align}\label{fml7}
	{{\bf{A}}_x} \!= \!\!\left[\!\!\! {\begin{array}{*{20}{c}}
	1& \!\!\!\!\!\cdots\!\!\!\!\! &1\\
	{{e^{ - j2\pi \!{f_k}d\cos \left( {{\theta _1}} \!\right)\sin \left( {{\varphi _1}} \!\right){\rm{ }}}}}& \!\!\!\!\!\cdots\!\!\!\!\! &{{e^{ - j2\pi \!{f_k}d\cos \left( {{\theta _I}} \!\right)\sin \left( {{\varphi _I}} \!\right)}}}\\
	\!\!\!\vdots\!\!\! & \!\!\!\vdots\!\!\! & \!\!\!\vdots\!\!\! \\
	{{e^{ - j2\pi \!{f_k}md\cos \left( {{\theta _1}} \!\right)\sin \left( {{\varphi _1}} \!\right)}}}& \!\!\!\!\!\cdots\!\!\!\!\! &{{e^{ - j2\pi \!{f_k}md\cos \left( {{\theta _I}} \!\right)\sin \left( {{\varphi _I}} \!\right)}}}
	\end{array}} \!\!\!\!\right]\!,
	\end{align}
	and 
	\begin{align}\label{fml8}
	{{\bf{A}}_y} \!= \!\!\left[\!\!\! {\begin{array}{*{20}{c}}
	1& \!\!\!\!\!\cdots\!\!\!\!\! &1\\
	{{e^{ - j2\pi \!{f_k}d\sin \left( {{\theta _1}} \!\right)\sin \left( {{\varphi _1}} \!\right)}}}& \!\!\!\!\!\cdots\!\!\!\!\! &{{e^{ - j2\pi \!{f_k}d\sin \left( {{\theta _I}} \!\right)\sin \left( {{\varphi _I}} \!\right)}}}\\
	\!\!\!\vdots\!\!\! & \!\!\!\vdots\!\!\! & \!\!\!\vdots\!\!\! \\
	{{e^{ - j2\pi \!{f_k}nd\sin \left( {{\theta _1}} \!\right)\sin \left( {{\varphi _1}} \!\right)}}}& \!\!\!\!\!\cdots\!\!\!\!\! &{{e^{ - j2\pi \!{f_k}nd\sin \left( {{\theta _I}} \!\right)\sin \left( {{\varphi _I}} \!\right)}}}
	\end{array}} \!\!\!\!\right],
	\end{align}
	respectively.
	Inspired by \cite{kotaru2015spotfi}, here, we take multiple subcarriers into consideration and extend the 2D AoA estimation into three dimensions, to acheive joint 2D AoA and ToF estimation via the following Proposition~\ref{L2}:
	\begin{prop}\label{L2}
	The signal 2D AoA and ToF at time $t$ can be estimated using
	\begin{align}\label{fml9}
	{P_{3D}}\left( {\theta ,\varphi ,\tau , t} \right){\rm{ = }}\frac{1}{{{\bf{A}}_{0x'y'}^{\rm{H}}{{\bf{E}}_N}\left( t\right){\bf{E}}_N^{\rm{H}}\left( t\right){{\bf{A}}_{0x'y'}}}},
	\end{align}
	where $P_{3D}$ describes the signal magnitude for a given set of $\left( {\theta ,\varphi ,\tau } \right)$, the superscript H is the conjugate transpose operator, ${{\bf{E}}_N}\left( t\right) $ is the noise subspace obtained by decomposing the auto-correlation matrix of the smoothed original signal at time $t$~\cite{kotaru2015spotfi}, ${{\bf{A}}_{0x'y'}}$ is the steering matrix that is obtained using~\eqref{fml7} and \eqref{fml8} as	
	\begin{align}
	&{{\bf{A}}_{0x'y'}} = {\left[ {{{\bf{A}}_0}{\rm{ }} {{\bf{A}}_{x'}}{\rm{ }}{{\bf{A}}_{y'}}} \right]^{\rm{T}}}
	\notag\\&
	=\!\! {\left[\! {\underbrace {\begin{array}{*{20}{c}}
	\!	1\! \\ 
	\!	\vdots \! \\ 
	\!	1 \!\\ 
	\!	\vdots \! \\ 
	\!	1 \!\\ 
	\!	\vdots \! \\ 
	\!	1 \!
	\end{array}}_{{{\mathbf{A}}_0}}\underbrace {\begin{array}{*{20}{c}}
	{{e^{ - j2\pi {f_1}d\cos \left( \theta  \right)\sin \left( \varphi  \right)}}} \\ 
	\vdots  \\ 
	{{e^{ - j2\pi {f_{{k'}}}d\cos \left( \theta  \right)\sin \left( \varphi  \right)}}} \\ 
	\vdots  \\ 
	\!\!	{{e^{ - \!j2\pi\!{f_1}\!\left(\! {{m'} \!- 1} \!\right)d\!\cos \left( \theta  \right)\!\sin \left( \!\varphi  \!\right)}}}\!\! \\ 
	\vdots  \\ 
	\!\!\!\!\!\!{{e^{ - \!j2\pi\!{f_{{k'}}}\!\left(\! {{m'}\! - 1} \!\right)d\!\cos \left( \theta  \right)\!\sin \left(\! \varphi  \!\right)}}}\!\!\! \!\!\!
	\end{array}}_{{{\mathbf{A}}_{x'}}}\underbrace {\begin{array}{*{20}{c}}
	{{e^{ - j2\pi {f_1}d\sin \left( \theta  \right)\sin \left( \varphi  \right)}}} \\ 
	\!\!\!\!\vdots\!\!\!\!  \\ 
	{{e^{ - j2\pi\!{f_{{k'}}}d\sin \left( \theta  \right)\sin \left( \varphi  \right)}}} \\ 
	\!\!\!\!\vdots\!\!\!\!  \\ 
	\!\!	{{e^{ - \!j2\pi\!{f_1}\!\left( \!{{n'}\! - 1} \!\right)d\!\sin \left( \theta  \right)\!\sin \left(\! \varphi \! \right)}}}\!\!\!\!\! \\ 
	\!\!\!\!\vdots \!\!\!\! \\ 
	\!\!\!\!\!\!{{e^{ -\! j2\pi\!{f_{{k'}}}\!\left( \!{{n'} \!- 1} \!\right)d\!\sin \left( \theta  \right)\!\sin \left( \!\varphi \! \right)}}} \!\!\!\!\!\!
	\end{array}}_{{{\mathbf{A}}_{y'}}}} \!\right]^{\text{T}}}\!,
	\end{align}
	$0 < {k'} < K$, $0 < {m'} < M$, and $0 < {n'} < N$.
	\end{prop}
	
	Thus, we complete all the processes of the inverse semantic-aware wireless sensing framework. Specifically, we use {\bf{Algorithm~\ref{Algorithm2}}} to sample the task-related signal spectrums. With the RIS, {\bf{Algorithm~\ref{Algorithm1}}} can encode the sensing data, thus greatly reducing the data volume to be stored or transmitted. We use the self-supervised decoding {\bf{Algorithm~\ref{Algorithm3}}} to recover the original sensing data. Finally, with the help of Proposition~\ref{L2}, various sensing tasks can be performed. For example, intrusion detection can be achieved by detecting the change of estimated 2D AoA, and the human walking trajectory can be tracked by estimating the 2D AoA and the ToF of the signals.
	
	\section{Experiments Results}\label{SS6}
	%
	Since the key contribution of this paper is to achieve the inverse semantic-aware encoding and decoding of the sensing data with the help of RIS, we aim to answer the following research questions via experiments:
	\begin{enumerate}
	\item[{\textbf{Q1)}}] Can the proposed self-supervised decoding scheme recover the original signal spectrums and ensure the accomplishment of sensing tasks?
	\item[{\textbf{Q2)}}] Can the amplitude response matrix of RIS, i.e., codebook, encrypt the sensing data?
	\item[{\textbf{Q3)}}] Compared with the existing uniform sampling method, can the proposed semantic hash sampling method help to achieve more accurate completions of the sensing tasks?
	\end{enumerate} 
	We first present the experimental platform and the parameter setting of our proposed algorithms, and then answer the above questions through experimental evaluations.
	
	\subsection{Experiments Setting}
	\begin{figure}[t]
	\centering
	\includegraphics[width=0.41\textwidth]{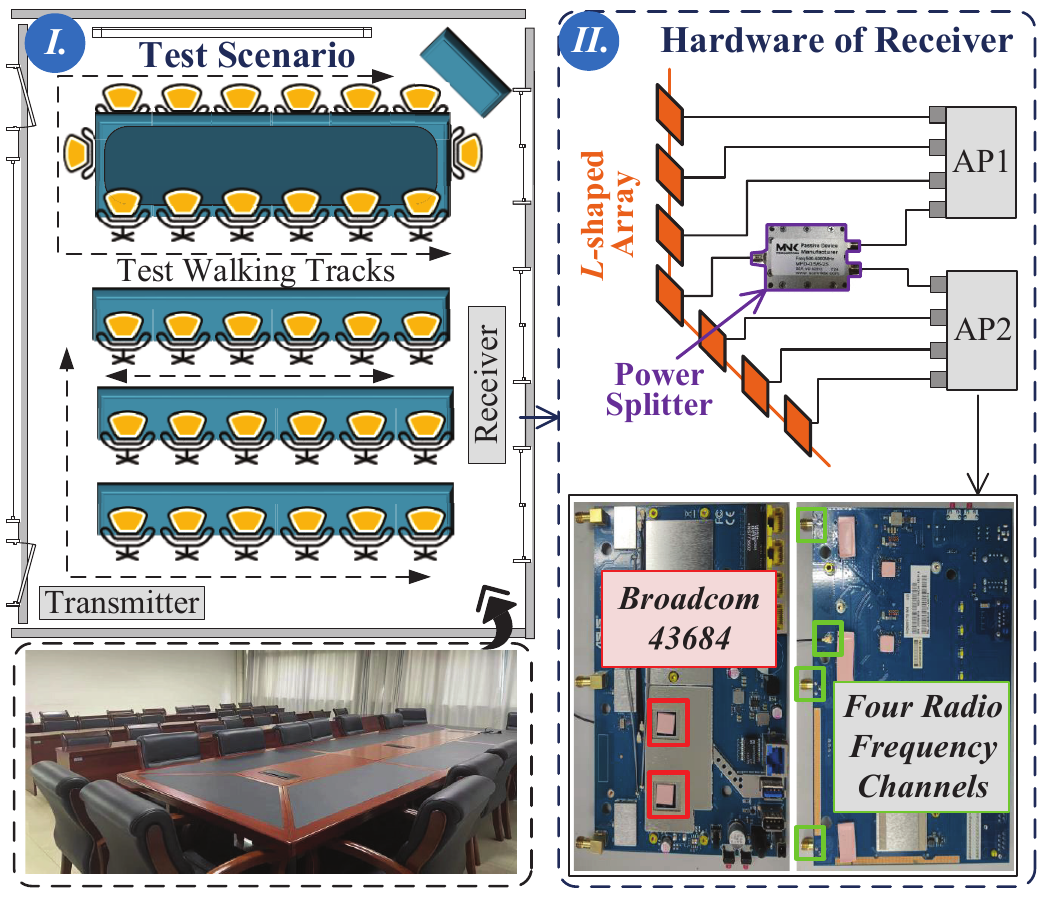}
	\caption{Test scenario and hardware of the receiver.}
	\label{experiment}
	\end{figure}
	To collect sensing data from the real-world scenario, we use three access points (APs) based on the IEEE 802.11ax protocol to build a test platform~\cite{gringoli2022ax}. The collected sensing data is used to conduct a comprehensive evaluation of our proposed algorithms. The specific experimental scenario and hardware equipment are shown in Fig.~\ref{experiment}. Specifically, the test scenario is a conference room with tables and chairs. Inside the room, one AP acts as a transmitter to send OFDM wireless signals with the total bandwidth of $160$ MHz and $2048$ sub-carriers. The center frequency of the sub-carriers is $5.805$ GHz. As shown in Fig.~\ref{experiment} (Part II), the other two APs form an receiver with $L$-shaped active sensor array via a power splitter to receive signals. Since the investigation of STAR-RIS hardware is still at a very early stage, we simulate the amplitude and phase response matrices of the transmissive elements using a signal processor~\cite{tang2022transmissive,mu2021simultaneously,xu2021star}. During the experiment, the data packet transmission rate, i.e., transmission frequency, is $100$ Hz, which means $100$ packets are transmitted per second. The human target walks along the preset trajectory to complete the data collection.
	
	The experimental platform for running our proposed algorithms is built on a generic Ubuntu 20.04 system with an AMD Ryzen Threadripper PRO 3975WX 32-Cores CPU and an NVIDIA RTX A5000 GPU. In the self-supervised decoding {\bf{Algorithm~\ref{Algorithm3}}}, two U-net without the skip connections~\cite{ulyanov2018deep} are used as the self-supervised neural networks, i.e., ${{\bm{\mathcal T}}\!_{{{\bm{\Theta }}_A}}}\!\!\left( {\bm{e}} \right)$ and ${{\bm{\mathcal T}}\!_{{{\bm{\Theta }}_P}}}\!\!\left( {\bm{e}} \right)$. The input to the network, i.e., ${\bm e}$, is a random vector that has the same size as ${\bm{x}}_A$ and ${\bm{x}}_P$ to be recovered. During the decoding of one MetaSpectrum, ${\bm e}$ is fixed in each ADMM iteration. In addition, to avoid the local minimum that the networks stuck in the last iteration, ${{\bm{\Theta }}_A}$ and ${{\bm{\Theta }}_P}$ are set to zero when each ADMM iteration is finished. In other words, both ${{\bm{\Theta }}_A}$ and ${{\bm{\Theta }}_P}$ are re-trained in each iteration.
	
	\subsection{Experiments Performance Analysis}
	\subsubsection{Effectiveness and Efficiency of the proposed inverse semantic decoding method (Q1)}
	We first set the data compression ratio as $10\%$. As shown in Fig.~\ref{res} (Part I), starting from $3$ seconds, we select one pair of amplitude and phase spectrums in each $0.1$ second time segment by using {\bf{Algorithm~\ref{Algorithm2}}}, for RIS-aided encoding. Using the encoding {\bf{Algorithm~\ref{Algorithm1}}} presented in Section~\ref{CM}, we can obtain one amplitude MetaSpectrum and one phase MetaSpectrum as shown in Fig.~\ref{res} (Part III) for every $10$ pairs of signal spectrums. The decoded results after $15$ iterations of the outer loop are shown in Fig.~\ref{res} (Part II). For both amplitude and phase spectrums, we observe that the difference between the decoded and the original spectrums is basically negligible. We present a detailed comparison of the decoded and the original amplitude spectrums in Fig.~\ref{res} (Part IV). This proves the effectiveness of our encoding and decoding methods.
	
	\begin{figure*}[t]
	\centering
	\includegraphics[width=0.9\textwidth]{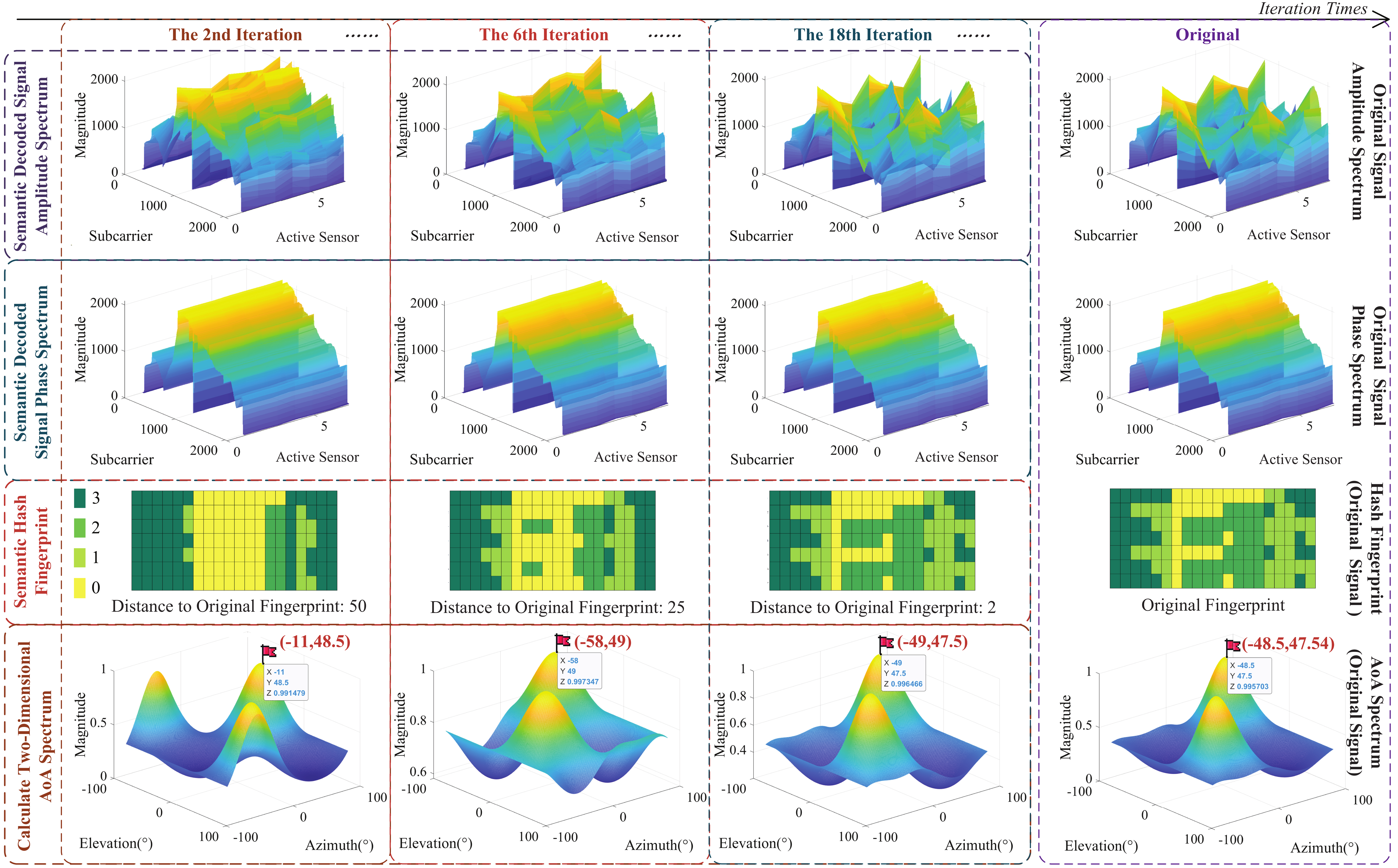} 
	\caption{The variation of the decoded amplitude and phase spectrums at time $4.5$ s in the experiment, the corresponding semantic hash matrix, the estimated 2D AoA spectrum by Proposition~\ref{L2} with the number of outer loop decoding iterations, where the data compression ratio is $5\%$.}
	\label{gradchange}
	\end{figure*}
	In addition to the visual contrast, we show in Fig.~\ref{gradchange} how the proposed semantic hash matrix changes with the number of outer loop decoding iterations. We set the data compression ratio as $5\%$. We observe that, as the number of outer loop iterations increases, both the decoded amplitude and phase spectrums at time $4.5$ seconds are gradually close to the ground truth spectrums. Moreover, the Hamming distance between the semantic hash matrices of the decoded pair of amplitude and phase spectrums and that of the original signal spectrums is gradually reducing. Specifically, we can see that $12$ iterations can make the Hamming distance only $2$, which takes about $40$ seconds on average. Furthermore, the estimated 2D AoA values using the decoded spectrum after $12$ iterations are very close to the true values, which basically has no effect on the practical sensing tasks. This proves the efficiency of our encoding and decoding methods.
	
	\subsubsection{Effectiveness of using the amplitude response matrix of the RIS as the codebook (Q2)}
	\begin{figure}[t]
	\centering
	\includegraphics[width=0.47\textwidth]{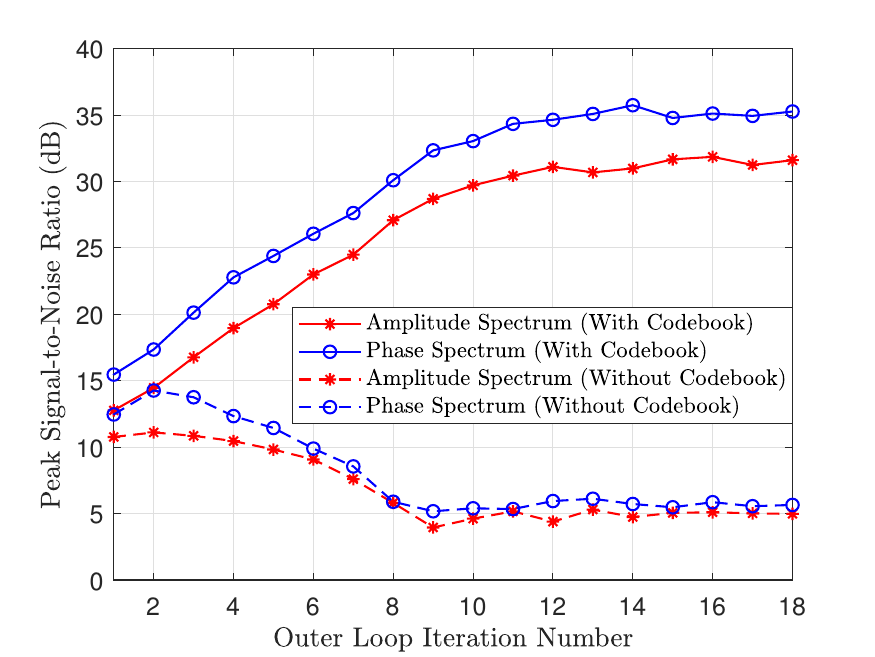}
	\caption{The PSNR values versus the number of outer loop decoding iterations with or without codebook.} 
	\label{iter}
	\end{figure}
	Figure~\ref{iter} depicts the average peak signal-to-noise ratio (PSNR) values $10$ experiments versus the number of outer loop decoding iterations, with or without the codebook ${\bf \Phi}_{A}^{\left(i\right)}$. If the codebook is available, {\color{black}we observe that the PSNR values of both the amplitude and phase spectrums are increasing as the number of iterations increases, and gradually reach a plateau after about $10$ iterations. Although minor fluctuations occur at higher iteration steps because of the dynamic nature of the decoding process, the overall trend demonstrates the effectiveness of our self-supervised decoding scheme in recovering the original signal spectrums.} However, if no codebook is available or the codebook is wrong, the PSNR values decrease as the number of iterations increases. The reason is that the parameters of two decoding network, i.e., ${{\bm{\Theta }}_A}$ and ${{\bm{\Theta }}_P}$, are learned according to a wrong objective function.

	\subsubsection{Effectiveness of the proposed semantic hash sampling method (Q3)}
	\begin{figure}[t]
	\centering
	\includegraphics[width=0.47\textwidth]{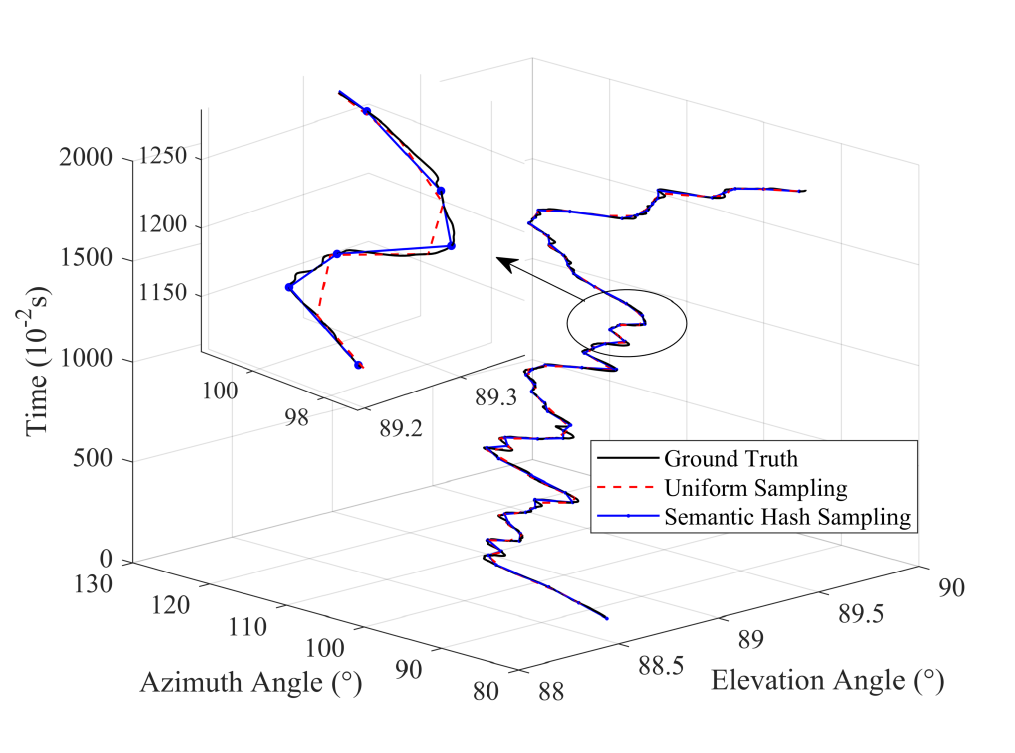} 
	\caption{Comparison between different sampling methods and ground truth in terms of 2D AoA changes with movement of human.}
	\label{trace}
	\end{figure}
	Based on the sensing data extracted via two different sampling methods, i.e., red line for the uniform sampling and blue line for the semantic hash sampling methods, Fig.~\ref{trace} displays estimated elevation and azimuth AoA changes over time. Note that the estimation results under two sampling schemes are obtained using the decoded amplitude and phase spectrums with the data compression ratio as $5\%$. First, we observe that the both elevation and azimuth AoA at every moment can be accurately estimated using the decoded data. This further validates the effectiveness of our proposed encoding and decoding algorithms (for Q1). Furthermore, by comparing the blue and red lines, it can be seen that the proposed semantic hash sampling method is more efficient and effective than uniform sampling in describing the details of AoA changes, as shown in the enlarged part in Fig.~\ref{trace}. Because these changes are typically more informative, this shows the effectiveness of our proposed semantic hash sampling method. To compare the two schemes numerically, we consider the MSE between the ground truth and the 2D AoA estimation results after interpolation. By calculating, we obtain that the estimation error of the semantic hash sampling is $0.89$, which is $67\%$ lower than that of uniform sampling scheme whose estimation error is $2.7$.
	
	\begin{figure}[t]
	\centering
	\includegraphics[width=0.47\textwidth]{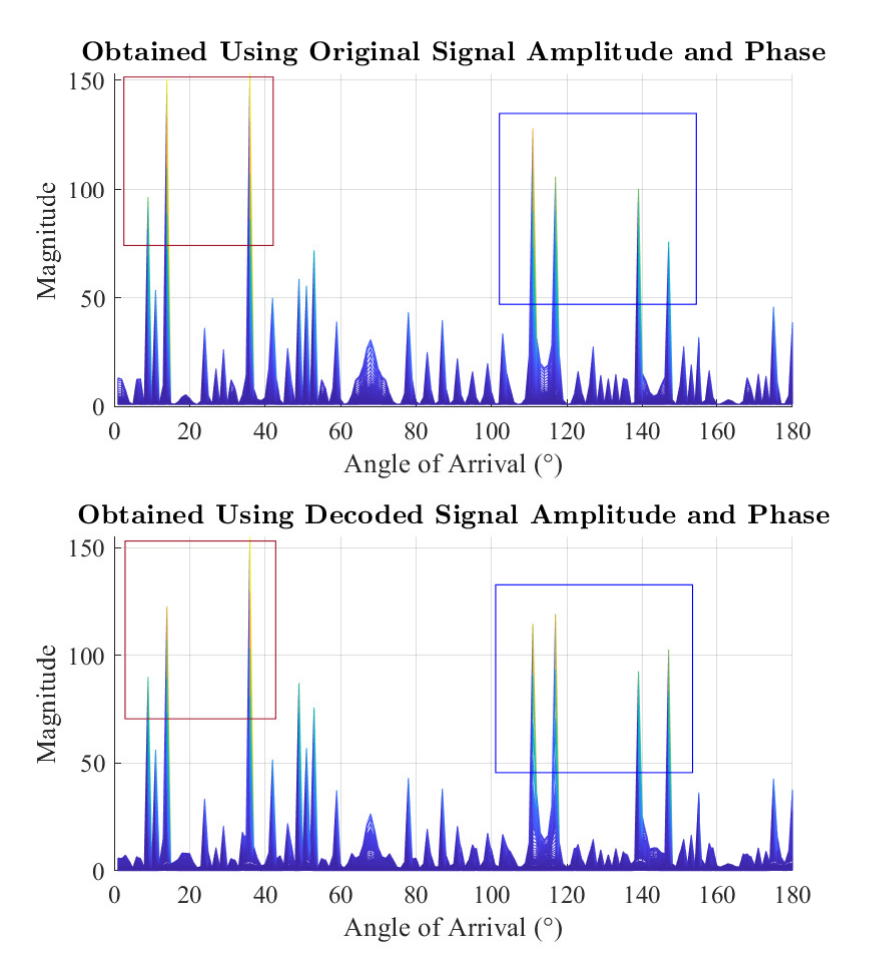} 
	\caption{Comparison of azimuth AoA estimation results that are obtained by using the original and decoded signals, respectively.}
	\label{SPFIG}
	\end{figure}
	In addition to the walking human, stationary objects such as tables and chairs in the conference room also reflect wireless signals. Thus, the information that can be extracted from the signal spectrums at one certain moment is rich. Taking the azimuth AoA as an example, Fig.~\ref{SPFIG} shows the comparison of the azimuth AoA estimation results that are obtained by using the original and decoded signals, respectively, with a data compression ratio of $5\%$. {\color{black}The results demonstrate that our encoding and decoding methods effectively preserve the semantic information related to the sensing tasks, which is evident from two aspects: First, the relative magnitude characteristics among different azimuth AoA estimated from the decoded signals are consistent with the ground truth, i.e., the azimuth AoA estimated from the original signals. For instance, the ground truth indicates that the stronger signals' azimuth AoA are in the range of $10^\circ - 40^\circ$ and $110^\circ - 150^\circ$, as marked by the red and blue boxes, respectively. In addition, the signals with AoA in $40^\circ - 110^\circ$ are weaker. These features are almost entirely preserved in the estimation results obtained using the decoded data. Second, we observe that the AoA estimation results of the first several strongest signals are almost unchanged before and after the inverse semantic-aware encoding and decoding, e.g., the signals marked by the red and blue boxes in Fig.~\ref{SPFIG}, respectively. This indicates that our proposed algorithms can effectively preserve the phase characteristics (for Q1).}
	
	\section{Conclusion and Future Directions}\label{SF}
	We have designed an inverse semantic-aware wireless sensing framework. The amplitude response matrix of the RIS can be effectively used to generate the codebook as prior knowledge for decoding. We have shown that our proposed RIS-aided encoding method can achieve effective data compression. When selecting the signal spectrums to be encoded, our proposed semantic hash sampling method is significantly better than the widely used uniform sampling method. Moreover, the self-supervised decoding method can recover signal amplitude and phase spectrums to achieve various wireless sensing tasks without affecting the performance. Since the decoding method does not require any pre-training, it can greatly save network resources. As the demand for sensing data increases, our proposed framework can contribute to building a resource-friendly next-generation Internet.
	\begin{figure*}[t]
	\centering
	\includegraphics[width=0.85\textwidth]{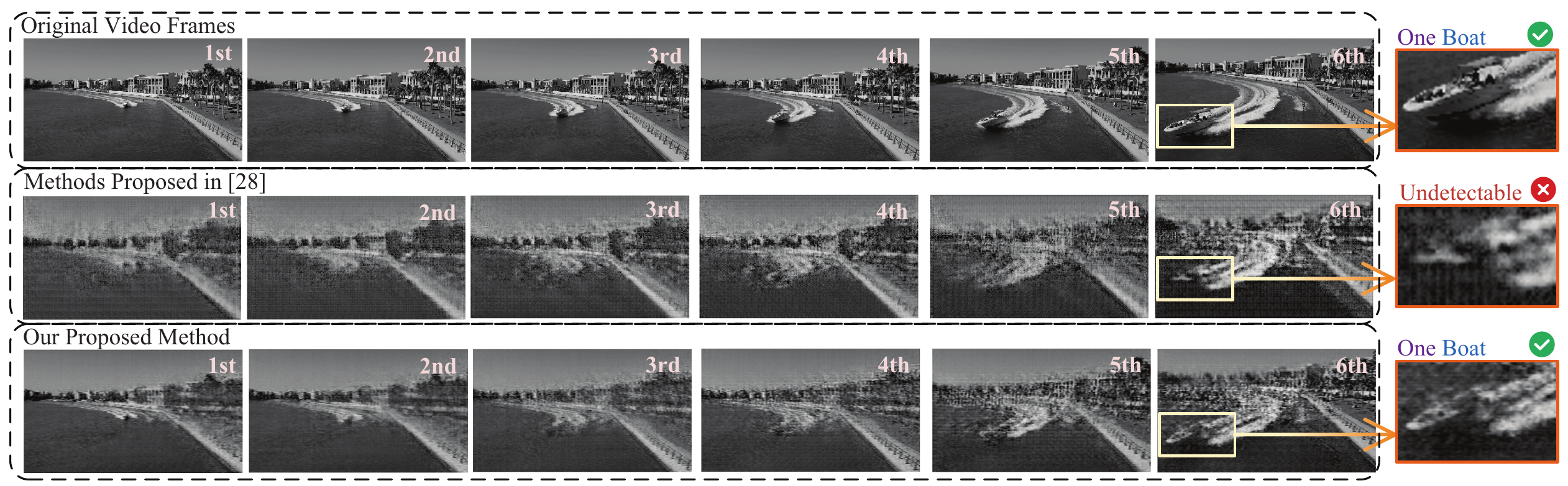} 
	\caption{Early-stop decompress results the real video frames in~\cite{perrin2020eyetrackuav2} using the method proposed in [25] and our method, respectively. Six images are compressed into one image.} 
	\label{realplot}
	\end{figure*}
	
	There are two potential future research directions.
	\begin{itemize}
	\item {\color{black}{\textit{Inverse Semantic-aware Transmission of Images}}. We can explore the application of inverse semantic-aware encoding and decoding for images or audio. In the context of surveillance services, for example, a camera captures a bay to detect boats. The surveillance videos require significant storage resources. The goal is to compress multiple video frames, such as the six frames depicted in Fig.~\ref{realplot}, into a single frame. The original frames can then be reconstructed using the proposed self-supervised decoding algorithm.}
	\item {\textit{Cantor or Szudzik Pairing Compression}}. In this paper, we encoded the amplitude and the phase spectrum separately. A possible improvement is to use the pairing functions, e.g., cantor~\cite{lisi2007some} or szudzik~\cite{szudzik2006elegant} pairing functions, to combine the two spectrum into one. As shown in Fig.~\ref{res}, the pairing compression can be used as an operation after obtaining amplitude and phase spectrums to further compress the sensing data.
	\end{itemize}

	\bibliographystyle{IEEEtran}
	\bibliography{IEEEabrv,Ref}
	\end{document}